%% file: main.tex
\documentclass[aps,prd,reprint,superscriptaddress,floatfix,nofootinbib,showkeys,showpacs,preprintnumbers]{revtex4-2}
\usepackage{tabularx}
\usepackage{xspace}
\usepackage{listings}
\usepackage{lineno}
\usepackage{bm}		%
\usepackage{amsmath}	%
\usepackage{amssymb}	%
\usepackage[normalem]{ulem}

\usepackage[dvipsnames,svgnames]{xcolor} 
\usepackage{graphicx}
\usepackage{comment}

\usepackage{natbib}
\usepackage{hyperref}
\usepackage{amsmath}
\usepackage{enumitem}
\setlist{wide, labelwidth=!, labelindent=0pt}
\usepackage{multirow}
\hypersetup{    
  colorlinks      = true,
  linkcolor       = {black},
  linkbordercolor = {white},
  citecolor       = {black},
  citebordercolor = {white},
  urlcolor        = {blue},
  urlbordercolor  = {white},
}
\usepackage{soul} 
\usepackage{amsmath}
\usepackage{amssymb}
\usepackage{xspace}
\usepackage{xifthen}

\mathchardef\mhyphen="2D
\newcommand{\vect}[1]{\boldsymbol{#1}}

\newlength{\dhatheight}

\newcommand{\bandvar}[2][]{%
  \ifthenelse{\isempty{#1}}{\var{#2}}{\var{#2\_#1}}%
}

\newcommand{\var}[1]{\ensuremath{\texttt{\MakeUppercase{#1}}}\xspace}

\newcommand{\omegam}{\Omega_{\rm{m}}}

\newcommand{\clustercomb}{CL+GC}
\newcommand{\allcomb}{CL+\ttt{}}
\newcommand{\ttt}{3$\times$2pt}

\defcitealias{Simpaper}{Simpaper}
\defcitealias{DESY1KP}{DESY1KP}
\defcitealias{DES_cluster_cosmology}{DESCL}

\providecommand\physrep{\ref@jnl{Phys.~Rep.}}%
\providecommand\apjs{\ref@jnl{ApJS}}%
\providecommand{\jcap}{\ref@jnl{JCAP}}%

\newcommand{\RNum}[1]{\uppercase\expandafter{\romannumeral #1\relax}}

\newcommand{\maglim}{\texttt{maglim}}
\newcommand{\mdet}{\texttt{Metadetection}}

 \usepackage{lipsum}
\usepackage{aas_macros}

\usepackage{tikz}
\usepackage{calc}
\def\checkmark{\tikz\fill[scale=0.4](0,.35) -- (.25,0) -- (1,.7) -- (.25,.15) -- cycle;} 
\newcommand{\redmapper}{{\rm redMaPPer}}

\begin{document}

\title{Dark Energy Survey: Modeling strategy for multiprobe cluster cosmology and validation for the Full Six-year Dataset }
\preprint{DES-2025-0883}
\preprint{FERMILAB-PUB-25-0114-PPD}

\include{authorsuper}

\collaboration{DES Collaboration}
\date{Affiliations at the end of the paper. Contact author: \url{chto@uchicago.edu}}
\begin{abstract}
We introduce an updated To$\&$Krause2021 model for joint analyses of cluster abundances and large-scale two-point correlations of weak lensing and galaxy and cluster clustering (termed \allcomb{} analysis)  and validate that this model meets the systematic accuracy requirements of analyses with the statistical precision of the final Dark Energy Survey (DES) Year 6 (Y6) dataset. The validation program consists of two distinct approaches, \emph{(i)} identification of modeling and parameterization choices and impact studies using simulated analyses with each possible model misspecification \emph{(ii)} end-to-end validation using mock catalogs from customized Cardinal simulations that incorporate realistic galaxy populations and DES-Y6-specific galaxy and cluster selection and photometric redshift modeling, which are the key observational systematics. In combination, these validation tests indicate that the model presented here meets the accuracy requirements of DES-Y6 for CL+3x2pt based on a large list of tests for known systematics. In addition, we also validate that the model is sufficient for several other data combinations: the CL+GC subset of this data vector (excluding galaxy--galaxy lensing and cosmic shear two-point statistics) and the CL+3x2pt+BAO+SN (combination of CL+3x2pt with the previously published Y6 DES baryonic acoustic oscillation and Y5 supernovae data).
\keywords{Cosmology, Cosmological parameters, Galaxy cluster counts,  Large-scale structure of the universe}
\pacs{98.80.-k, 98.80.Es, 98.65.-r}
\end{abstract}
\maketitle 
\maketitle 
\section{Introduction}
 Wide-field imaging surveys have been one of the primary tools to study cosmic structure and its time evolution. Recent studies from the Stage-\RNum{3} surveys, such as the Dark Energy Survey (DES) \citep{Y3kp},  Hyper Suprime Cam Subaru Strategic Project (HSC-SSP) \citep{HSC2}, and Kilo-Degree Survey (KiDS) \citep{KIDS3x2pt2}, have presented exquisite measurements of the amplitude of the cosmic structure fluctuations. However, much of the signal-to-noise in these measurements lies in the regime where complex measurement, observational, and modeling systematics hinder the interpretation. To deliver unbiased and competitive cosmological constraints from these measurements, one must judiciously choose the range of the data vector used for the analyses and the complexity of the model so that the accuracy of the analysis matches the statistical precision of the measurement. 

Achieving this analysis precision becomes even more challenging when combining multiple cosmological probes. In particular, DES is designed to enable measurements of four key probes of dark energy: galaxy cluster abundances, galaxy clustering, weak gravitational lensing, and type-Ia supernovae \citep{2005astro.ph.10346T}. Each probe is sensitive to different aspects of cosmic structure, and their combination breaks degeneracies between cosmological and systematic parameters, leading to more accurate constraints. The higher accuracy of the measurement places a stronger requirement on the precision of the analysis. In addition, as astrophysical and observational systematics affect each cosmological probe differently, establishing a self-consistent model across multiple probes becomes particularly challenging.

This paper outlines the program for joint analyses of galaxy cluster abundances, galaxy clustering, weak gravitational lensing, baryon acoustic oscillation (BAO), and type-Ia supernovae for the final Dark Energy Survey year 6 (DES-Y6) dataset\footnote{See appendix \ref{app:listofmodel} for a comprehensive list of improvements from DES-Y1 analysis \citep{4x2pt1, 4x2pt2}.}. The combination of galaxy cluster abundances, galaxy clustering, and weak gravitational lensing involves six two-point correlation functions and is referred to as \allcomb{} hereafter.
The major part of the paper focuses on the analysis involving galaxy clusters and their cross-correlations with galaxies and lensing (\clustercomb{}), which includes cluster abundances, cluster--galaxy cross-correlations, cluster clustering, galaxy clustering, and cluster lensing. This is because type-Ia supernovae and BAO are expected to have uncorrelated systematics, and the modeling of the joint analyses of galaxy clustering and weak gravitational lensing for the same data set has been studied extensively \cite{Y6model}. The cluster-focused data vector follows the DES-Y1 analysis \citep{4x2pt1}, incorporating cluster–galaxy cross-correlations, cluster clustering, and galaxy clustering to constrain cluster mass via the halo bias–halo mass relation. Cluster lensing provides an independent mass constraint. The combined data vectors constrain cluster mass and self-calibrate cluster selection effects. The constrained cluster mass and cluster abundances then lead to competitive cosmological constraints. We further note that this paper pays particular attention to cluster-specific systematics. While some of the systematics that impact the \ttt{} analysis are also tested for the combination of the probes, we do not attempt to exhaustly test all possible \ttt{} systematics. In this work, we adopt two different approaches to validate our analysis program and demonstrate that the accuracy of our program matches the expected precision of the measurements from the DES-Y6 data. The first approach is to theoretically calculate possible modeling systematics and quantify the impact of each systematic on our cosmological constraint. The second approach is to use numerical simulations to provide an alternative modeling test and to validate the combination of modeling systematics, astrophysical systematics, and sample selections. We detail these two approaches below. 

We first explicitly calculate various hypothetical modeling and observational systematics. We then validate the amplitude and direction of each individual systematic's impact on our cosmological constraints. This approach, known as the simulated likelihood analysis, has been adopted as a primary tool to test the precision of the models in survey analyses \citep{Y6model,y3method, DESmethod}.

While the first approach is particularly powerful in understanding each individual systematic's impact on cosmological constraints, it has one important caveat. Several modeling and astrophysical systematics are hard to calculate analytically, such as the cluster selection function, the performance of photometric redshift estimations, and the galaxy selection function. This is particularly problematic for cluster analyses. In the analysis of galaxy clusters, the major challenge lies in understanding the selection function and its impact on cluster abundances and related two-point correlation functions, including cluster lensing, cluster--galaxy cross-correlations, and cluster clustering. Without a proper model of this selection, cosmological constraints can be strongly biased \citep{DES_cluster_cosmology}. 

Previous studies have investigated the possible origin of the selection bias. These include the impacts of the projected line-of-sight structure, often referred to as the projection effect, on the stacked cluster lensing profile ($\Delta \Sigma (r)$) \cite{Tomomi_projection, Andres2024, Heidiselection}, how this projection effect can affect the cluster abundances \citep{Matteoprojection, Andres2024}, and how the triaxial shape of the halos combined with a preferential selection of certain orientations can affect weak lensing profiles \cite{2023MNRAS.523.1994Z,2019MNRAS.490.4889H}. While promising, these studies often focus on a specific hypothesized physical origin that affects the cluster selection, only investigate a specific part of the theory model (e.g., $\Delta \Sigma (r)$), and mostly rely on approximate cluster-finding algorithms that may not mimic the \redmapper{} selection in detail. 

In this paper, while we also ensure that our model is flexible enough that the previously investigated causes of non-trivial cluster selection functions will not impact our cosmological constraints, we also use a realistic simulation, Cardinal \citep{cardinal}, in which \redmapper{} is run in the same setting as run on real data, to provide additional validation of the cluster selection and its impact on all the considered data vectors, including cluster abundances, cluster clustering, cluster--galaxy cross-correlations, and cluster lensing. We note that this validation is on the real observables (e.g., tangential correlation function $\gamma_t (\theta)$) instead of the intermediate theory model (e.g., projected matter density profile $\Sigma (r)$). This is enabled by running photometric redshift estimation algorithms and DES-like galaxy selection on the simulated catalog so that we can construct realistic galaxy and shape catalogs that incorporate various possible correlated systematics \citep{Y3Buzzard, Buzzardy1}. As such, analyses in Cardinal validate the pipeline from catalog to cosmology.  

This paper is organized as follows. In section~\ref{sec:model}, we detail the theoretical model for the observables. We begin with a general description of the theory model and proceed with the specific implementation (\ref{ssec:parameterizations}).  
In section~\ref{sec:mockuniverse}, we describe the construction of simulated DES observations and the processes of transforming those observations into the data vector.
In section~\ref{sec:simulated}, we employ the simulated likelihood analysis techniques to justify the implementation of the theory model described in  \ref{ssec:parameterizations}. 
In section~\ref{sec:mocktest}, we detail the result of validating the theory model using the data vector generated in simulated DES analyses. 
Section~\ref{sec:conc} provides the conclusion.

\section{Model}
\label{sec:model}
Our theory model and covariance calculation are implemented in \textsc{CosmoLike} \citep{cosmolike2016}. 
To identify potential sources of systematic errors in the theoretical modeling, we separate the theory model prediction into the general theoretical formalism for calculating angular \allcomb{} statistics (Sect.~\ref{sec:formalism},\ref{sec:summary}) and specific, often empirical, model parameterization choices (Sect.~\ref{ssec:parameterizations}), which may be subject to model misspecification. This setup allows us to systematically identify potential sources of systematic biases and validate each of our parameterization choices.

We follow the notation in the DES-Y3 method paper \citep{DESmethod}. In short, we use the lower-case Italic subscripts ($i$, $j$) for tomographic bin indices, lower-case Greek subscripts ($\alpha$) for the vector and shear components, lower-case Roman superscripts for specific samples (s: source galaxy, l: lens galaxy, c: cluster sample), lower-case Italic subscripts $p$ for a generic sample, $p \in (\rm{l, s, c})$, upper-case Italic subscripts to denote cluster richness bins, ${\rm{c}}_A$, and upper-case cursive subscripts 
$\mathcal{A} \in (\delta_{\rm{l}}, \delta_{\rm{c}_A}, \rm{E}/\rm{B})$ for projected fields.

\subsection{Projected tracer fields}
\label{sec:formalism}
\subsubsection{Lensing field}
The observed galaxy shapes are a spin-2 field with components $\gamma_{\alpha}$. 
Its two components $\gamma_\alpha$ are modeled as gravitational shear ($\mathrm G$) and intrinsic ellipticity. The latter is split into a spatially-coherent contribution from intrinsic galaxy alignments (IA), and stochastic shape noise $\epsilon_0$:
 \begin{equation}
 \gamma_{\alpha}^{j}(\hat{\mathbf n}) = \gamma_{\alpha,\mathrm G}^{j}(\hat{\mathbf n})+\gamma_{\alpha,\mathrm{IA}}^{j}(\hat{\mathbf n})+\epsilon_{\alpha,0}^j(\hat{\mathbf n})\,.
 \end{equation}
Shape noise contributes to the covariance but not to the mean two-point correlation function signal, hence we do not include it in the mean model prediction and refer to \citep{y3-cosmicshear1,y3-cosmicshear2,y3-covariances} for details on the covariance modeling.

At leading order, the gravitational shear field is related to the convergence field $\kappa_{\rm{s}}$, which for the $i$-th tomography bin of a generic sample $p$ can be calculated as 
\begin{equation}
\kappa_{p}^i (\hat n)=  \int d\chi W_{\kappa, p}^i(\chi)\delta_m^{(\rm{3D})}(\hat{n}\chi, \chi)\,.\\
\end{equation}
Here $\chi$ is the comoving distance, $\delta_m^{(\rm{3D})}$ is the three-dimensional matter density contrast, and $W^i_\kappa$ is the tomographic lens efficiency
\begin{equation}
W_{\kappa, p}^i(\chi)  =  \frac{3\Omega_mH_0^2}{2}\int_\chi^\infty d\chi^\prime n_{p}^{i} (\chi^\prime) \frac{\chi^\prime}{a(\chi)} \frac{\chi^\prime-\chi}{\chi^\prime}\,,
\end{equation}
with $n^i_p(z)$ is the redshift distribution of sample $p$ in the tomographic bin $i$.

The intrinsic alignment contribution to the observed galaxy shear field is a projection of the 3D field $\tilde{\gamma}_\mathrm{IA}$ 
 weighted by the source galaxy redshift distribution
\begin{equation}
 \gamma_{\alpha,\mathrm{IA}}^i(\hat{\mathbf n}) = \int d \chi \, W_{\delta,\rm{s}}^i(\chi) \tilde{\gamma}_{\alpha,\mathrm{IA}}\left(\hat{\mathbf n} \chi, \chi\right)\,.
\end{equation}
The 3D intrinsic alignment field, $\tilde{\gamma}_{\mathrm{IA}}$, is specified by the choice of intrinsic alignment model, c.f. Sect.~\ref{ssec:parameterizations}. 

Given the precision of DES-Y3 analyses, next-to-leading order corrections to the gravitational shear are negligible \citep{DESmethod}; hence, gravitational shear only produces an $\rm{E}$-mode component. However, next-to-leading order intrinsic alignment effects are relevant at the accuracy of our analysis and may produce both $\rm{E}$- and $\rm{B}$-modes.

\subsubsection{Galaxy and cluster density field}
In general, the observed projected density field ($\delta^i_{p} (\hat n)$) at position $\hat n$ of a generic sample in a given tomographic bin ($i$) includes contributions from projection of three-dimensional density contrasts ($\delta_{p,D}^i (\hat n)$), contribution from redshift space distortion ($\delta_{p, \rm{RSD}}^i (\hat n) $) that moves the samples in and out of the given tomographic bin, and magnification due to line-of-sight structures ($\delta_{p, \mu}^i (\hat n)$),
\begin{equation}
\delta_{p}^i (\hat n) = \delta_{p,\rm{D}}^i (\hat n)+ \delta_{p,\rm{RSD}}^i(\hat n) +\delta_{p,\mu}^i(\hat n)\,.
\end{equation}
The projection of three-dimensional density contrasts can be calculated as 
\begin{eqnarray}
    \delta_{p,D}^i (\hat n) &=& \int d\chi W_{\delta, p}^i(\chi)\delta_p^{(\rm{3D})} (\hat{n}\chi, \chi),  \\
    W_{\delta, p}^i(\chi)  &=&  n^i_p(z) \frac{dz}{d\chi}, 
\end{eqnarray}
where $\chi$ is the comoving distance, $\delta_p^{\rm{3D}}$ is the three-dimensional density contrast, and $n^i_d(z)$ is the redshift distribution of the generic samples in the tomographic bin $i$. The contribution from redshift space distortion ($\delta_{p, \rm{RSD}}^i (\hat n) $) can be modeled as 
\begin{eqnarray}
    \delta_{p, \rm{RSD}}^i (\hat n) &=& -\int d\chi W_{\delta, p}^i(\chi) \frac{\partial}{\partial \chi} \left (\frac{\hat{n} \vect{v} (\hat{n} \chi, \chi)}{a(\chi)H(\chi)}\right), 
\end{eqnarray}
where $\vect{v}$ is the peculiar velocity of the generic sample, $a$ is the scale factor, and $H(\chi)$ is the Hubble rate. Note that in the above equation, we have assumed that $\vect{v}$ is small compared to the width of the tomographic bin. Since we only focus on large scales for this analysis, we further relate $\hat{n} \vect{v}$ to the density contrast of the samples via the linearized continuity equation. Finally, the magnification caused by line-of-sight structures can be modeled as 
\begin{eqnarray}
   \delta_{p, \mu}^i (\hat n) &=& C_p^i \kappa_p^i(\hat{n}), %
  \label{eq:magnification}
\end{eqnarray}
where $C_p$ is a magnification bias coefficient.

\subsection{Field to summary statistics}
\label{sec:summary}
\subsubsection{Two-point statistics}
\label{ssec:2pt}
With the exception of the angular galaxy/cluster clustering power spectra $C^{ii}_{\delta_{\rm{l/c}_A}\delta_{\rm{l/c}_A}} (\ell)$, which are evaluated including non-Limber contributions \citep{Fang_nonlimber}, we calculate the angular cross-power spectrum between two projected fields $\mathcal{A,B}$ using the Limber approximation  \begin{align}
     C_{\mathcal{AB}}^{ij}(\ell) = \int d\chi \frac{W_{\mathcal{A}}^i(\chi)W_{\mathcal{B}}^j(\chi)}{\chi^2}P_{\mathcal{AB}}\left(k = \frac{\ell+0.5}{\chi},z(\chi)\right)\,,
     \label{eq:CAB}
 \end{align}
with $P_{\mathcal{AB}}$ the corresponding three-dimensional power spectrum, which is specified by the model choices detailed in Sect.~\ref{ssec:parameterizations}.

The angular two-point statistics in configuration space are computed from the corresponding angular power spectrum using the curved-sky transforms
\begin{align}
    w^{ij}_{p p'}(\theta) =& \sum_\ell \frac{2\ell+1}{4\pi}P_\ell(\cos\theta) C^{ij}_{\delta_{p}\delta_{p'}}(\ell)~,\label{eq:transform_w}\\ \label{eq:ggl}
     \gamma_{t,p}^{ij}(\theta) =& \sum_\ell \frac{2\ell+1}{4\pi\ell(\ell+1)}P^2_\ell(\cos\theta) C^{ij}_{\delta_{p}\mathrm{E}}(\ell)~,\\
     \nonumber \xi_{\pm}^{ij}(\theta) =& \sum_\ell\frac{2\ell+1}{2\pi\ell^2(\ell+1)^2}[G_{\ell,2}^+(\cos\theta)\pm G_{\ell,2}^-(\cos\theta)]\\
     &\times \left[ C^{ij}_{EE}(\ell)\pm C^{ij}_{BB}(\ell)\right]~,\label{eq:transform_xi} \end{align}where $P_\ell$ and $P_\ell^2$ are the Legendre polynomials and the associated Legendre polynomials, $G_{\ell,m}^{+/-}$ are given by Eq.~(4.19) of \cite{1996astro.ph..9149S}.
We calculate the correlation functions within an angular bin $[\theta_{\rm min},\theta_{\rm max}]$ by carrying out the angular bin average of the transformation kernels analytically \citep{y3-covariances}.

For galaxy--galaxy lensing, we add the tangential shear contribution of an enclosed mass to the $\gamma_t(\theta)$ prediction to remove the one-halo contribution in the galaxy--galaxy lensing data vector at large scales. This process is known as the point-mass marginalization \citep{MacCrann_2019}. Specifically, we add the following term to equation \ref{eq:ggl}, 
\begin{eqnarray}
   \nonumber \Delta \gamma_{t,g}^{ij}(\theta) = \frac{1}{\delta_{\rm crit}\omegam} \int d\chi W^i_{\delta, g} (\chi) W^j_{\kappa, s} (\chi)\frac{B^i 10^{15} M_{\odot}/h}{d_A(\chi)^2\theta^2}, \\
\label{eq:PM}
\end{eqnarray}
where $B^i$ are free parameters that we marginalize over and $d_A(\chi)$ is the angular diameter distance.

For cluster lensing, we further transform the tangential shear profile $\gamma_{t,\rm{c}}$ into the projected surface density 
\begin{equation}
\label{temaki}
\Sigma_{\rm{c}_A}^{ij}(\boldsymbol{\theta}) = \mathbf{Y} \gamma_{t,\rm{c}_A}^{ij} (\boldsymbol{\theta}),
\end{equation} 
with $\mathbf{Y}$ a tranformation matrix defined in \citep{Ytransform}. While the tangential shear profile includes non-local contributions from all enclosed mass, the projected surface density at a given $\theta$ only probes the cluster density profile within the corresponding cylindrical annulus and thus removes contributions of the small-scale cluster mass profile to the large-scale cluster lensing signal. 

\subsubsection{Cluster abundances}
The number of galaxy clusters in richness bin $A$ in cluster tomography bin $\delta z_c^i$ can be calculated as
\begin{align}
\label{eq:hm}
\nonumber    N^i_A =&\int_0^{\infty} d z_{\rm{true}} \frac{dV}{dz_{\rm{true}}}\int_{z_{\rm{obs}}\in \delta z_c^i} p(z_{\rm{obs}}|z_{\rm{true}})\\
    &\int_{\lambda\in A} \!\!\!\!d\lambda \int_{0}^{\infty} dM\, p(\lambda|M,z_{\rm{true}}) \frac{dn}{dM}(M,z_{\rm{true}})\,,
\end{align}
with $dV/dz_{\rm{true}}$ the survey volume per unit redshift, $p(\lambda|M,z_{\rm{true}})$ the richness--mass relation, and $dn/dM$ the halo mass function. Our parameterization choices for the latter two model ingredients are specified in Sect.~\ref{ssec:parameterizations}.
\subsection{Model parameterization choices}
\label{ssec:parameterizations}
\subsubsection{Matter power spectrum}
To compute lensing (cross-) power spectra, we relate the convergence to 3D density contrast using Eq.~\ref{eq:CAB} and
\begin{equation}
 P_{\kappa \mathcal{A}}(k,z) =  P_{\mathrm{m} \mathcal{A}}(k,z)\,.
 \label{eq:kappaX}
\end{equation}

For DES-Y6 analyses, we adopt the halo model-based HMcode2020 \citep{HMcode2020} as non-linear matter power spectrum $P_{\rm{mm}}(k,z)$, including baryonic feedback with $T_{\rm{AGN}} = 7.7$  as a fixed parameter. %

We refer to \citep{Y6model} for validation of this model choice in the context of the 3$\times$2pt analysis and demonstrate the robustness of the \allcomb{} to baryon feedback model and galaxy bias model misspecification in Sect.~\ref{ssec:Pmm_test}.

\subsubsection{Halo Mass and Halo Bias}
We adopt the Tinker fitting functions for the halo mass function $\frac{dn}{dM}(M,z)$ and linear halo bias $b_{1,\rm{h}}(M,z)$ \citep{Tinker10}. The robustness of our analyses to model misspecifications in halo mass function and halo bias prescriptions is validated in Sect.~\ref{ssec:emu_test}.

\subsubsection{Richness--Mass Relation}
Following DES-Y1 \citep{4x2pt1}, we model the richness--mass relation ($p(\lambda | M, z_{\rm{true}})$) as a log-normal model with scatter
\begin{equation}
   \sigma_{\rm{ln}\lambda}^2 = \sigma_{\rm{intrinsic}}^2+(e^{\langle \rm{ln}\lambda \rangle}-1)/e^{2\langle \rm{ln}\lambda \rangle}, 
\end{equation}
 and mean
 \begin{equation}
 \langle \ln(\lambda) |M \rangle = \rm{ln}\lambda_0 + A_{\lambda} \ln (M/M_{\rm{piv}}) + B_{\lambda} \ln \left (
 \frac{1+z}{1.45}\right).
 \end{equation}

\subsubsection{Tracer Bias} 
On scales corresponding to the two-halo regime, we adopt a linear bias ($b_1$) prescription relative to the \emph{nonlinear} matter density as the baseline model, such that the cross power spectrum between galaxy/cluster density and field $\mathcal{A}$ is given by
 \begin{equation}
 P_{\delta_{p} \mathcal{A}}(k,z) = b_{1,p}(z) P_{\mathrm{m} \mathcal{A}}(k,z)\,.
 \label{eq:PgX}
 \end{equation}
For galaxies, we model the redshift dependence with one free parameter $b_{1,g}^i$ per tomographic bin per tracer sample, neglecting the evolution of galaxy bias within tomographic bins. 

For galaxy clusters, the redshift evolution of $b_{1,\rm{c}}^i$ is calculated from the redshift evolution of the halo mass and halo bias function, 
\begin{align}
\label{eq:bias}
\nonumber    b_{1,\rm{c}_{A}}^i(z) =&\Bigg[\int_{\lambda\in A} \!\!\!\!d\lambda \int dM\, p(\lambda|M,z) \frac{dn}{dM}(M,z) b_{1,\rm{h}}(M,z)\Bigg]\\
&\times \Bigg[\int_{\lambda\in A} \!\!\!\!d\lambda \int dM\, p(\lambda|M,z) \frac{dn}{dM}(M,z)\Bigg]^{-1}\,.
\end{align}
Following the 3$\times$2pt methodology, contributions from non-linear galaxy biasing are mitigated by scale cuts\footnote{We note that this is one of the two DES-Y6  \ttt{} analysis approaches.}. We demonstrate the robustness of the linear bias model for the \allcomb{} analysis in Sect.~\ref{ssec:NL_test} (galaxy bias) and Appendix~\ref{app:NL} (cluster halo bias).

Following DES-Y1 \citep{4x2pt1,cosmolike2016}, we also add the contribution of the one-halo term to the cluster lensing power spectrum. Specifically, $P_{\delta_{c} \mathcal{A}}(k,z)$ is modeled by an additional term written as 
\begin{align}
\nonumber   &P^{1h}_{\delta_{\rm{c}_A}\mathcal{A}}(k,z)=\\
&\Bigg[\int_{\lambda\in A} \!\!\!\!d\lambda \int dM\, p(\lambda|M,z) \frac{dn}{dM}(M,z) \frac{M}{\bar{\rho}_{\rm{m}}}u(k,c,z, M)\Bigg] \nonumber\\
&\times \Bigg[\int_{\lambda\in A} \!\!\!\!d\lambda \int dM\, p(\lambda|M,z) \frac{dn}{dM}(M,z)\Bigg]^{-1}\,
\end{align}
where $\bar{\rho}_{\rm{m}}$ is the mean matter density of the universe,
and $u(k,c,z, M)$ is the Fourier transform of the NFW profile with halo concentration $c$ and mass $M$, for which we use the concentration--mass relation of \cite{2013ApJ...766...32B}. Note that we do not include one-halo contributions for the modeling of the galaxy-galaxy lensing because those contributions are explicitly removed through the point-mass marginalization (Eq.~\ref{eq:PM}).

\subsubsection{Selection effect model}
\label{sec:selection_model}
The identification and the richness estimation of clusters can be affected by large-scale environments. Such a correlation manifests as an additional bias for cluster-related correlation functions. Here, different from the scale-independent model adopted in DES-Y1 \citep{4x2pt1}, we find it necessary to include the scale-dependency in the selection model due to the lower scale cut in our analysis program (section \ref{sec:datavec}) compared to DES-Y1. Specifically, we find that analyzing the contaminated data vector with a scale-independent selection effect model results in a bias in $S_8$ and $\omegam$ constraints by $0.45$ of the statistical uncertainty.  Our model is motivated by HOD-cylinder mocks generated with methods described in \citep{2024arXiv241002497L}. The functional form reads, 

\begin{eqnarray}
\label{eq:sel}
  \frac{\Sigma_{\rm{c}_A}^{ij}}{\Sigma_{\rm{c}_A}^{ij} [\rm{Orig}] } (\theta) &=&  b_{s1} + b_{s2} \rm{exp}\left(-\frac{\theta\chi(\bar{z})}{r_0}\right) \\
  \frac{\Sigma_{\rm{c}_A}^{ij}}{\Sigma_{\rm{c}_A}^{ij} [\rm{Orig}] } (\theta) &=&\frac{w^{ii}_{\rm{c}_A\rm{g}}}{w^{ii}_{\rm{c}_A\rm{g}} [\rm{Orig}] }(\theta)\nonumber
\\
   \frac{\Sigma_{\rm{c}_A}^{ij}}{\Sigma_{\rm{c}_A}^{ij} [\rm{Orig}] } (\theta)&=& \sqrt{\frac{w^{ii}_{\rm{c}_A\rm{c}_A}}{w^{ii}_{\rm{c}_A, \rm{c}_A} [\rm{Orig}] }(\theta)}, \nonumber
\end{eqnarray}
where $\bar{z}$ is the mean redshift of the cluster sample in redshift bin $i$, $\Sigma_{\rm{c}_A}^{ij}  [\rm{Orig}] $ is the original cluster lensing model described in equation \ref{temaki}, $w^{ii}_{\rm{c}_A\rm{g}} [\rm{Orig}]$  and $w^{ii}_{\rm{c}_A, \rm{c}_A} [\rm{Orig}]$ are described in equation \ref{eq:transform_w}, and $b_{s1}$, $b_{s2}$, $r_0$ are free parameters. The first equation of equation \ref{eq:sel} is motivated by HOD-cylinder mocks in appendix \ref{app:HOD-cylindermock}, where we find that on large scales, the additional modulation of the cluster--matter correlations follows a simple power law. Given that the projected over-density is proportional to cluster-matter correlations, we expect their ratio to follow the same functional form. A similar trend in the projected over-density is also found in alternative mocks \cite{Heidiselection, Sunayama2023}. Our parametrized model can be easily understood. $b_{s1}$ controls the large-scale asymptotic behavior while $b_{s1}+b_{s2}$ controls the amplitude at zero separation. $r_0$ controls the transition scale. The second equation of equation \ref{eq:sel} is expected due to our assumption that galaxy density is a linear matter density tracer (equation \ref{eq:PgX}). The third equation can be understood as follows. The additional modulation in the matter overdensity around clusters is due to cluster selection. One can attribute this modulation to the cluster density field, which contributes twice to the cluster clustering signal leading to a quadratic boost. This quadratic boost is validated in  HOD-cylinder mocks presented in appendix \ref{app:HOD-cylindermock}.

\subsubsection{Intrinsic Alignments}
We adopt the ``tidal alignment and tidal torquing'' (TATT) model \citep{TATT} as the baseline intrinsic alignment model for the analyses presented here, similar to various DES analyses \citep{y3-cosmicshear2,Y3kp, Y6model}.
Briefly, the intrinsic galaxy shape field is written as an expansion in the density and tidal tensor $s$,
\begin{align}
\tilde{\gamma}_{\alpha,\mathrm{IA}} = A_1 s_{\alpha} +A_{1\delta} \delta_\mathrm{m} s_{\alpha} + A_2 \left(s\times s\right)_{\alpha} + \cdots\,.
\label{eq:TATT}
\end{align}
The linear term, with $A_1$, corresponds to the well-studied ``nonlinear linear alignment'' model \citep[NLA,][]{Catelan2001,Hirata_IA,Bridle_NLA}. The second term captures the impact of source density weighting
\citep{Blazek15}, and together, the two comprise the `tidal alignment' component. The third term, quadratic in the tidal field, captures the impact of tidal torquing \citep{Catelan2001,lee2001}.

The DES-Y6 baseline model adopts a four-parameter parameterization of the IA amplitude and redshift dependence, fixing $A_{1\delta}$ to $A_1$, and choosing to parameterize the redshift evolution of $A_1$ and $A_2$ as power laws with exponents $\eta_{1,2}$. In detail, the prefactors are given by
\begin{align}
A_1(z) =& -a_1\bar{C}_1\frac{\rho_\mathrm{crit}\Omega_{\rm m}}{D(z)}\left(\frac{1+z}{1+z_0}\right)^{\eta_1}\\
A_2(z) =& 5a_2\bar{C}_1\frac{\rho_\mathrm{crit}\Omega_{\rm m}}{D(z)^2}\left(\frac{1+z}{1+z_0}\right)^{\eta_2} ,
\end{align}
with pivot redshift $z_0$ corresponding to the mean redshift of the source sample\footnote{We note that the mean redshift corresponds to the mean redshift of the source sample in DES-Y1. We keep this value the same for ease of comparison.}, $\bar{C}_1$ a normalization constant, which by convention is fixed to $\bar{C}_1 = 5\times10^{-14} M_\odot h^{-2}\mathrm{Mpc}^2$, and $D(z)$ the linear growth factor.%

We note that the DES-Y3 baseline model includes the density weighting term with a redshift-independent amplitude $b_{\rm{TA}}$ (which is directly proportional to $A_{1\delta}$ in Eq.~\ref{eq:TATT}), yielding a five-parameter IA model. For the cluster lensing data vector, we further reduce the model to $A_2=0$ because we find that intrinsic alignment has negligible impact on cluster lensing.  Specifically, comparing the data vector generated from the upper and lower $1\sigma$ values of $A_1$ constrained from DES-Y3, the changes in $\chi^2$ for cluster lensing is $0.04$.

\subsubsection{Magnification}
Magnification modifies the observed galaxy density through geometric dilution and the modulation of galaxy flux and size selection \citep{VernerV1995, Moessner1998,sizebias}. Hence the DES Y3/Y6 3$\times$2pt analyses model the lensing bias coefficient $C_{\rm{l}}^i$ introduced in Eq.~\ref{eq:magnification} as 
\begin{equation}
C_{\rm{l}}^i = 5 \frac{\partial \ln n_{\rm{l}}^i}{\partial m}\bigg\rvert_{m_\mathrm{lim},r_\mathrm{lim}} +\frac{\partial \ln n_{\rm{l}}^i}{\partial \ln r}\bigg\rvert_{m_\mathrm{lim},r_\mathrm{lim}} -2,
\end{equation}
where the logarithmic derivatives are the slope of the galaxy magnitude and size distribution at the sample selection limit.

The values of galaxy lensing bias coefficients $C_{\rm{l}}^i$ are estimated from simulations and data \cite{y3-2x2ptmagnification,Y6Mag}. In Y6 analyses, we marginalize over uncertainties of the magnification bias estimates, as validated in \citep{Y6Mag}. For Y3 cosmology analyses, it was sufficient to hold magnification bias parameters fixed, as demonstrated in validation on mock catalogs \citep{Y3Buzzard}.

The observed cluster density is only affected by geometric dilution,
\begin{equation}
C_{\rm{c}_A}^i \equiv -2\,,
\end{equation}
as we absorb the magnification-induced modulation of observed cluster richness in the mass--richness relation parameterization.

\subsubsection{Photometric redshift uncertainties}
In this paper, we follow DES-Y3 models of photometric redshift uncertainties. Specifically, we adopt an additive shift parameter for the mean redshift for each tomographic bin for source galaxies, reading as,  
\begin{eqnarray}
    n_s^i(z) = n^i_{\rm ref} (z-\Delta_z^i).
\end{eqnarray}
For the lens galaxies, we adopt a stretch parameter to change the width of the redshift distribution in addition to a shift parameter in each tomographic bin. Mathematically, this is expressed as 

\begin{equation}
  n^i(z) = \frac{1}{\sigma_z^i} n^i_{\rm ref} \left(
  \frac{z-\langle z \rangle}{\sigma_z^i}+\langle z \rangle \right).
\end{equation}

\subsection{Covariance matrix}
\label{sec:cov}
The DES-Y3 3$\times$2pt analysis relies on analytic covariances, which schematically consist of Gaussian and non-Gaussian cosmic variance, weak lensing shape noise, (Poisson) clustering shot noise, and super-sample covariance. These analytic covariances are calculated using the halo model-based \textsc{CosmoCov} code \citep{kre17,CosmoCov} and validated in \citep{y3-covariances}.
In DES-Y1 \citep{2021MNRAS.502.4093T}, we extended the DES 3$\times$2pt analytic covariances to include cluster abundances and configuration-space cluster-density cross-correlations. 

Relative to DES-Y1 \citep{2021MNRAS.502.4093T}, the analysis choice presented in this paper incorporates two significant updates: (1) Following the Y3 3$\times$2pt covariance validation \citep{y3-covariances}, we implement the exact curved-sky, angular bin-averaged transformation from angular power spectrum covariance to configuration-space statistics for all \allcomb{} statistics. (2) We transform the covariances for two-point correlations of $w,\, \gamma_{t},\xi_{\pm}$ type provided by 
\textsc{CosmoCov} into a covariance for $\Sigma_{\rm{c}}$ and its cross-covariance with other observables: we apply the same localizing transform to the covariance as to the data vector. Schematically, for a subset of the data vector consisting of $\Sigma^{ij}_c(\boldsymbol{\theta})$ and a type of observable $x$ 
\begin{equation}
\mathrm{Cov}_{\Sigma_{\rm{c}},x} = \begin{bmatrix} \mathbf{Y} &0\\ 0 &1 \end{bmatrix} \mathrm{Cov}_{\gamma_{t,\rm{c}},x} \begin{bmatrix} \mathbf{Y} &0\\ 0 &1 \end{bmatrix}^t\,.
\end{equation}
\section{Mock universe generation}
\label{sec:mockuniverse}
We validate our theory using the Cardinal simulation \citep{cardinal}. In this section, we summarize the construction of the Cardinal simulation, emphasizing the differences between the simulation used and those presented in \citep{cardinal}. We then describe the construction of the source, lens, cluster catalogs, and the associated DES analysis products necessary for the cosmological analyses, including the photometric redshifts, systematic weights, and magnification biases. Finally, we summarize the construction of the \allcomb{} data vectors and the associated priors for the cosmological analysis. Our analysis philosophy is similar to that of previous DES analyses \citep{6x2pt+NMethod, Y3Buzzard}. We treat the simulation as a plausible universe, where we run data analysis pipelines in a similar way as they were run on real data. We then validate our ability to recover the true cosmology in the simulated universe. Effectively, we are validating our pipeline starting from existing catalogs similar to the DES Gold catalog \citep{Y6Gold} and shear catalog \citep{Y6Shear} to cosmological constraints. A more complete end-to-end would also include pixel-level images, but that is beyond the scope of this paper.

\subsubsection{Cardinal simulation}
\begin{figure*}
    \centering
    \includegraphics[width=0.9\textwidth]{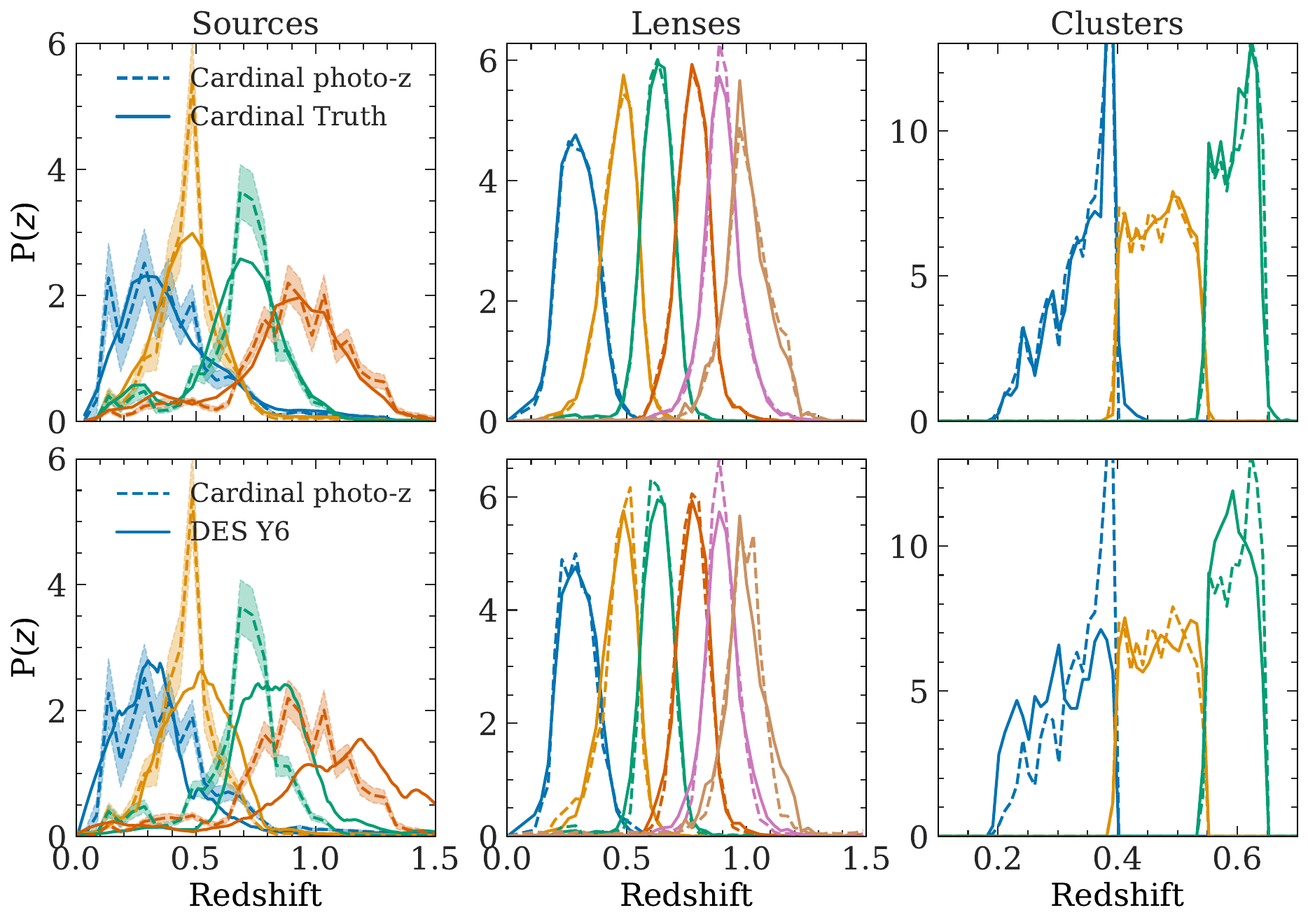}
    \caption{Comparison of redshift distributions between photometric redshift estimations in Cardinal, true redshift distributions in Cardinal (solid lines; top), and DES-Y6 data (solid lines; bottom). The first column shows source galaxies, the second column shows \maglim{} galaxies and the third column shows galaxy clusters. }
    \label{fig:zdist}
\end{figure*}
The Cardinal simulation is the successor of the Buzzard simulation \citep{Buzzard}, which has been heavily used for DES cosmological analyses \citep[e.g.][]{Buzzardy1, Y3Buzzard}. Cardinal and Buzzard both consist of galaxies with an r-band magnitude $m_{\rm r}<27$ spanning $10$k $\rm{deg}^2$ of the sky with $z=0$--$2.35$. The simulated catalogs were constructed from three N-body simulations with sizes of $1.05^3$, $2.6^3$, $4.0^3$ $h^{-3}\rm{Gpc}^3$ and mass resolution of $3.3\times 10^{10}$, $1.6\times 10^{11}$, and $5.9\times 10^{11}$ $h^{-1}M_\odot$. The N-body simulation is generated using \textsc{L-GADGET2} \citep{gadget} with a $\Lambda$CDM cosmology with $\omegam=0.286, \Omega_{\rm b}=0.047, \sigma_8=0.82, n_{\rm{s}}=0.96, h=0.7, A_s=2.145\times 10^{-9}$, and three massless neutrino species with $\rm{N}_{eff}=3.046$. These three N-body simulations are combined to generate a $10$k $\rm{deg}^2$ N-body lightcone, on which we paint galaxies. 

Galaxies are painted through the following steps. First, we assign galaxy positions, velocities, and absolute magnitudes to dark-matter particles based on their Lagrangian overdensities. The relation of galaxy properties and the Lagrangian overdensities is obtained from a subhalo abundance matching model constrained by the galaxy--galaxy correlations and group--galaxy correlations measured in SDSS \citep{SDSS}. This process was detailed in \citep{addgals}. The main difference between Cardinal and its predecessors \citep{JoeBuzzard, Y3Buzzard} is the generation of the subhalo abundance matching model. Cardinal is the first of its kind that includes orphan subhalos whose disruptions are constrained by the group--galaxy and galaxy--galaxy correlations. This additional modeling component significantly improves the fidelity of galaxy clustering below $2\ h^{-1}{\rm Mpc}$ (see Fig. 5 of \cite{cardinal}), which is essential for reproducing cluster abundances in simulations. Second, each galaxy is assigned an SDSS SED \citep{SDSS} using a conditional abundance matching algorithm. To efficiently assign galaxy colors, we further simplify the SDSS SED as a linear combination of five templates generated by the Kcorrect algorithm \citep{kcorrect03, kcorrect07}.  Compared to the predecessors, we modify the property used for conditional abundance matching so that it does not create an artificial color gradient in halos with a mass greater than $M_{200b}=10^{13}\ h^{-1}M_\odot$. Third, we generate the broadband DES $ugrizY$ and VISTA $JHK$broadband magnitudes by integrating the SEDs with the corresponding bandpass filters. In principle, one could apply DES-Y6 noise on these broadband magnitudes to obtain a galaxy catalog that mimics the DES-Y6 gold catalog, which was the procedure adopted in predecessors \citep{Y3Buzzard}. In practice, we find that the broadband galaxy colors generated with the procedure above assuming a DES-Y3 noise model are too red compared to the DES-Y3 data due to the limited diversity of SDSS SEDs and the insufficiency of describing galaxy colors as linear combinations of five SED templates. We address this problem with an iterative approach \cite{cardinal}. In short, we use the procedure above to generate a photometric catalog using a DES-Y3 noise model. We then correct the colors in this photometric catalog by comparing them to DES-Y3 data. We use the corrected photometric catalog to generate an SED library, which is then used to create a new photometric catalog. This catalog is then processed with a footprint-dependent DES-Y6 noise model to mimic the DES-Y6 gold catalog \citep{Y6Gold}. We refer the reader to section 3.6 of \citep{cardinal} for more details. In principle, one would like to re-tune Cardinal with the DES-Y6 gold catalog. However, we found that the Y3 calibrated version with the Y6 noise model can reproduce the overall magnitude distribution of DES-Y6 gold catalog at $10\%$ level (Fig. \ref{app:magcomp}), which is slightly worse than Cardinal presented in \citep{cardinal} but better than Buzzard \citep{Y3Buzzard}. We, therefore, decided not to retune the Cardinal model. %

We then perform multiplane ray-tracing \citep{calclens} to the mock galaxy catalog to compute each galaxy's shear, magnification, and deflection. Specifically, these quantities are computed using the lensing Jacobian at the equally placed shells from $z=0$ to $z=2.3$ with a separation $25\ h^{-1} \rm{Mpc}$. To speed up the process, each shell is divided into healpix pixels with $n_{\rm side}=8192$ (corresponding to a resolution of $0.46$ arcmin), where Jacobian lensing is computed. Although multiplane ray-tracing produces high-fidelity lensing quantities, it suffers from limited resolution where the lensing Jacobian is computed \cite{JoeBuzzard}. Specifically, the lensing signal at scales below $\sim10$ times the resolution biases low by over $10\%$ for cosmic shear and galaxy--galaxy lensing \citep{cardinal}. Cardinal is the first of its kind to address this resolution problem using a hybrid approach. Specifically, the shear of each galaxy is corrected according to the differences between $\Delta \Sigma$ computed using particle--halo cross-correlations and ray-tracing shears. We refer the reader to section 3.5 of \cite{cardinal} for details. We find that this hybrid approach can reduce the resolution effect from $10\%$ to less than $2\%$ at scales between $0.46$ arcmins (ray-tracing resolution) and $4.6$ arcmins for galaxies at $z=0.6$.%

Finally, we rotate the mock catalog into the DES-Y6 footprint, apply the DES-Y6 masks, and add footprint-dependent photometric noise according to algorithms detailed in \cite{cardinal}. Specifically, we assume the photometric noise is dominated by Poisson noise due to galaxy photons and sky background. We compute this noise using the effective exposure time and $10\sigma$ limiting magnitude maps in the DES-Y6 catalogs. Note that Cardinal is the first of the simulation series to adopt the Poisson model, while its predecessors approximate this Poisson process by assuming a Gaussian distribution. 

\subsubsection{DES Lens Catalog}
In this work, we use the \maglim{} \citep{2022PhRvD.106j3530P} sample as the lens galaxy sample, which is selected based on the Directional Neighbourhood Fitting \citep[DNF,][]{DNF} photometric redshifts. Following the implementation in DES, for each galaxy, DNF finds the $80$ nearest spectroscopic galaxies according to a directional metric, which is a product of the Euclidean metric and the angular metric, computed from $g,r, i,z$ magnitude. The point estimate redshift $z_{\rm DNF}$ is then given by a linear interpolation of redshifts of those $80$ nearest spectroscopic galaxies. DNF also provides the redshift corresponding to the nearest spectroscopic galaxies ($z_{\rm nn}$) for estimating the redshift distribution of an ensemble of \maglim{} galaxies. 

To generate DNF redshifts in Cardinal, we first generate mock spectroscopic training samples for DNF by selecting simulated galaxies with similar $i$-band magnitude and redshift distributions to the spectroscopic training samples for DNF in the DES-Y6 data \citep{Y6Lens}. We then run the DNF algorithm for all galaxies in Cardinal using the mock spectroscopic samples. With $z_{\rm DNF}$ in Cardinal, one could then select \maglim{} galaxies in Cardinal using the same magnitude cut as the DES-Y6 data \citep{Y6Lens}, namely, $m_i<18+4z_{\rm DNF}$. In practice, due to the differences between galaxy colors in Cardinal and DES-Y6 data, the \maglim{} generated with the same magnitude cut will have a different number density, leading to different signal-to-noise ratios between simulations and DES-Y6 data. Since the primary purpose of Cardinal is to validate our modeling assumptions, matching the signal-to-noise ratio between simulations and data is more important than adopting exactly the same magnitude cut. We, therefore, choose to modify the magnitude cut to minimize the number density differences between Cardinal and DES-Y6 \maglim{} samples. Specifically, we bin the Cardinal \maglim{} galaxies into six tomographic bins based on their $z_{\rm DNF}$ and bin edges $[0.2, 0.4 , 0.55, 0.70, 0.85 , 0.95, 1.05]$. We then fit a piecewise linear relation as the magnitude cut so that the difference in the \maglim{} number density of Cardinal and DES-Y6 data is minimized in each tomographic bin. The resulting changes in the magnitude cut are similar to DES-Y3 \citep{2022PhRvD.106j3530P}, ranging from $0.01$ mag in the lowest tomographic bin to $0.35$ in the highest tomographic bin. In the middle panel of figure \ref{fig:zdist}, we show a comparison of the redshift distribution between Cardinal DNF photometric redshift, Cardinal true redshift, and DNF photometric redshift of DES-Y6 \maglim{} samples. Here, we estimate the photometric redshift distribution using the DNF $z_{\rm nn}$. We find in Cardinal that the DNF $z_{\rm nn}$ photometric redshift distribution matches the true redshift distribution well and, therefore, adopts this estimation for cosmological analyses in Cardinal. We note that in the DES-Y6 data, a self-organizing map-based (SOM) photometric redshift estimation is used to determine the redshift distribution of \maglim{} galaxies, but validating that procedure in Cardinal is beyond the scope of this paper. 

The footprint-dependent photometric noise in Cardinal can cause spurious clustering signals on \maglim{} galaxies. To mitigate this, we weigh each \maglim{} galaxy based on weights generated based on the effective exposure time and $10\sigma$ limiting magnitude maps that are used to generate photometric noise in Cardinal. Here, we briefly describe the adopted procedure and refer the readers to \cite{Y6Lens} for details. We wish to remove the spurious clustering signal caused by the footprint-dependent survey systematics. Naively, one can perform a linear regression with L2 norm given all the maps describing survey systematics (survey property maps hereafter) and the observed galaxy overdensity maps. The weight of each galaxy is then the inverse of the predicted galaxy density based on the linear combination of survey property maps with coefficients given by the best-fit model. However, the large number of possible survey property maps might lead to overfitting and can remove real clustering signals in the galaxy overdensity maps. To avoid this, one can add regularization terms and determine their values via the cross-validation method \citep{enet}. Specifically, we choose the regularization terms as a linear combination of L1 and L2 norm, which has been demonstrated to perform well when a large number of coefficients tend to be zero \citep{enet, enet2}. The combined log-likelihood, also known as ElasticNet (or \textsc{ENET} hereafter), is then given by
\begin{equation}
\label{eq:salmon}
    \mathcal{L(\bm{\alpha})} = \frac{1}{2N_{\rm pix}} || \bm{\delta}_{\rm gal} - \bm{S} \bm{\alpha} ||_2^2-\lambda_1 ||\bm{\alpha}||_1 - \frac{\lambda_2}{2}||\bm{\alpha}||_2^2, 
\end{equation}
where $\bm{\alpha}$ are free parameters, $\bm{S}$ are survey property maps, $\lambda_{1,2}$ are hyperparameters, $N_{\rm pix}$ is the number of entries in $\bm{S}$, and $\bm{\delta}_{\rm gal}$ is the galaxy overdensity maps.  $\bm{\alpha}$, $\lambda_{1}$ and $\lambda_{2}$ are determined via cross-validation. The weight of each galaxy is then $(1+\bm{S} \bm{\alpha})^{-1}$. So far, we mostly follow the procedure described in \citep{enet}. In Cardinal, since the number of relevant survey property maps is much smaller than the real data, \textsc{ENET} might fail due to the fact that the distribution of $\bm{\alpha}$ can be very different from the real data. To avoid this, we consider all the survey property maps in the DES-Y6 data in addition to the relevant property maps when evaluating equation \ref{eq:salmon} even though the galaxy number density in Cardinal only depends on a small number of survey property maps. 

Finally, the galaxy overdensity is also modulated by the lensing of the foreground structure. Specifically, the galaxies behind an overdense region will be brighter and are more likely to pass the \maglim{} selection criteria. Moreover, their positions will change and can, on average, be less clustered due to lensing deflections. As detailed in section \ref{sec:model}, we model this effect by linearly relating galaxy overdensities to the tomographic convergence field $\kappa$. The coefficient of this relation is determined by Balrog simulations in DES-Y6 analyses. In Cardinal, we empirically measure this coefficient. Specifically, we select \maglim{} samples based on unmagnified magnitude. We then compare the fractional differences in the number of galaxies between magnified and unmagnified \maglim{} samples. We fit a linear model to the fractional differences given the $\kappa$ values, and the slope of the linear model is the magnification coefficient. %

\subsubsection{DES Source Catalog}
We construct a galaxy sample that mimics the DES-Y6 \mdet{} sample as the source catalog. The \mdet{} algorithm uses the code {\sc Ngmix}\footnote{\url{https://github.com/esheldon/ngmix}} to fit galaxy shapes and uses different artificially sheared images to self-calibrate the shear estimator. The DES-Y6 source catalog is then selected based on \mdet{} quantities such as signal-to-noise and size. 
Since Cardinal only has flux matching to the photometry in the \texttt{Gold} galaxy catalog \citep{Y6Gold}, we generate a different set of \mdet{}-specific galaxy properties for all Cardinal galaxies, which is then used for source galaxy selection. Here, we rely on the \texttt{Balrog} imaging simulations \citep{Y6Balrog, y3-balrog}, which inject low-noise galaxy images measured in the DES deep fields \citep{y6Deepfield} into wide-field images and remeasure their photometry through the DES Data Management pipeline. Assuming that the noise in deep field galaxy properties is negligible, we can use the \texttt{Balrog} simulation as a map from true galaxy properties to DES galaxy properties. Specifically, we find the \texttt{Balrog} injected galaxy closest in $griz$ bands for each Cardinal galaxy. We then apply the magnitude differences between deep and wide photometry of the matched \texttt{Balrog} galaxy to the corresponding Cardinal galaxy. The galaxy flux in Cardinal generated with this process will be referred to as $f^{\rm bal}$ for the rest of the paper. In principle, one would want to use the Y6 Balrog for this process. However, at the time of the Cardinal construction, Y6 Balrog did not exist. We, therefore, apply the following shortcut. Given a Y3 Balrog galaxy, we replace the noisy photometry with the \mdet{} galaxies whose corresponding $r, i,z$ magnitudes in the gold catalog are the closest. We find that the resulting modified Y3 Balrog has a similar magnitude and magnitude error distribution to the Y6 \mdet{} catalogs. We then use this modified Y3 Balrog to generate Cardinal source galaxies.

While using \texttt{Balrog} to generate noisy galaxy photometry in the simulation ensures a reasonable distribution of photometric noise given the true photometry, it has an important caveat. The generated photometric noise will not be realistically footprint-dependent, making it hard to use simulations to study the impact of source clustering due to survey nonuniformity. An important improvement of Cardinal over Buzzard is that we include footprint-dependent noise in $f^{\rm bal}$ so that the source galaxies and their redshift distributions in Cardinal will naturally inherit this footprint-dependent noise effect. Specifically, we use the fact that the \texttt{Gold} catalog photometry in Cardinal, albeit less accurate, has this footprint-dependent noise. We can then use the conditional abundance matching technique to shuffle the $f^{\rm bal}$ and $\sigma(f^{\rm bal})$ pairs so that $p(<\sigma(f^{\rm bal}) | f^{\rm bal}_i)=p(<\sigma(f) | f^{\rm bal}_i)$, where $\sigma(f)$ is the uncertainty of \texttt{Gold} catalog photometry.

We finally select the \mdet{} galaxies in Cardinal with the following cut: 
\begin{enumerate}
    \item $10<f^{\rm bal}_i/\sigma (f^{\rm bal}_i)<1000$
    \item $m_i<24$
    \item $m_g-m_z>0.7$
    \item $\sqrt{r_{50}^2+r_{\rm psf}^2}/r_{psf}>x_1/(1+x_2z)+x_3$, 
\end{enumerate}
where $r_{50}$ is the size of Cardinal galaxies and $r_{\rm psf}$ is the size of DES PSF at the position of each Cardinal galaxy. The first cut mimics the signal-to-noise ratio cut in the DES Y6 \mdet{} catalog \citep{Y6Shear}. The second and third cuts are to ensure reasonable photometric redshift uncertainties in Cardinal. The fourth cut has three parameters tuned to ensure that the number density of Cardinal is similar to that of DES-Y6 data. 

We then proceed to generate redshift bins and the corresponding redshift distribution for the selected source galaxy catalog. This process is done with the Self-Organizing Map Photometric Redshift (SOMPZ) algorithm \citep{sompz-y3}. The essential idea of the DES SOMPZ algorithm is to estimate the redshift distributions of an ensemble of galaxies by combining information of redshift samples and deep-field galaxy samples \citep{y6Deepfield}. To empirically estimate the redshift distributions with a limited number of redshift samples, we further reduce the dimensionality of the galaxies' color space into a finite number of categories (SOM cells) with a self-organizing map (SOM) algorithm. Mathematically, the SOMPZ redshift estimator can be expressed as a product of probability distributions, written as,   
\begin{align} 
    \label{eq:som_main}
     p(z|\hat{b},\hat{s}) 
     &\approx \sum_{\hat{c} \in \hat{b}} \sum_c p(z|c,\hat{s}) p(c|\hat{c}, \hat{s}) p(\hat{c}|\hat{s}), 
\end{align}
where $\hat{c}$ is the SOM cells of source galaxies,  $c$ is the SOM cells of deep-field galaxies, $\hat{b}$ denotes the tomographic bin, and $\hat{s}$ represents source galaxy selections. 

Each component in equation \ref{eq:som_main} is empirically evaluated in simulations in a manner similar to the data. Specifically, we first select patches in Cardinal that have the same area as the deep field in DES-Y6 data. In addition to the DES $ugriz$ photometry, we also generate Ultra Vista Survey $JHK$ photometry by integrating SEDs of each galaxy over the corresponding bandpass. We then remove the deep-field galaxies that will not pass the \mdet{} selection, which depends on the noisy photometry as described above. 
To make the \mdet{} selection of Cardinal deep-field galaxies, we generate noisy photometry by performing Monte-Carlo realizations on random locations across the DES footprint, following the same method used to create the source galaxy catalog in Cardinal. To mitigate noise from the Monte-Carlo process, we repeat this procedure ten times and apply the \mdet{} selection on each set of noisy photometry. Each deep-field galaxy is then assigned a weight ($w_i$) based on the number of times its noisy realizations meet the \mdet{} selection criteria. Subsequently, we apply a consistent level of photometric noise derived from the median depth of DES-Y6 data to each deep-field galaxy. This allows us to construct the deep SOM ($c$) based on realistic deep-field $ugrizJHK$ noisy photometry and properly take into account the effect of \mdet{} selections. To estimate the redshift distribution of the deep field galaxies ($p(z|c,\hat{s})$), we generate the simulated redshift samples by randomly selecting the same number of galaxies as the DES-Y6 redshift sample \citep{y6Redshift} out of the simulated deep-field galaxies. Here, we assume this redshift sample is free from selection biases. 

We construct the wide SOM ($\hat{c}$) using simulated \mdet{} source galaxies. We then construct the transfer function ($p(c|\hat{c}, \hat{s})$) from deep SOM ($c$) to wide SOM ($\hat{c}$) using the deep-field galaxies and their Monte-Carlo-realized noisy photometry described above. Specifically, we assign each Monte-Carlo-realized noisy photometric measurement to a wide SOM cell ($\hat{c}$) whose corresponding deep field photometry is also assigned to a deep SOM cell ($c$). We can then calculate the joint probability $p(c, \hat{c} | \hat{s})$ by calculating the fraction of the Monte Carlo-realized deep galaxy that is assigned to both $c$ and $\hat{c}$. Further, we can calculate $p(\hat{c} | \hat{s})$ by calculating the Monte Carlo-realized deep galaxy assigned to  $\hat{c}$. Using Bayes theorem, we get 
\begin{eqnarray}
    p(c|\hat{c}, \hat{s})=\frac{p(c, \hat{c} | \hat{s})}{p(\hat{c} | \hat{s})}, 
\end{eqnarray}. 

Finally, we assign each wide SOM cell $\hat{c}$ to a tomographic bin. To do so, we use the bin edge $z=[0.0, 0.404, 0.59, 0.82,2.0]$ to ensure a roughly equal number density of source galaxies per bin, similar to how the bin edge is chosen in DES-Y6 analyses. We then assign each redshift sample in Cardinal to a bin based on its redshift. For each wide SOM cell $\hat{c}$, we find the mode of the bin in which redshift samples corresponding to the wide SOM cell belong. We then assign this mode value as the bin associated with the wide SOM cell.

The photometric redshift generated above is noisy due to the limited number of spectroscopic samples and the limited size of the deep field. To reflect this noise, we employ the three-step Dirichlet sampling ($3s$DIR) method \citep{3sdir} to generate $10^4$ realizations of the redshift distribution $n(z)$ by analytically calculating the impact of shot noise and sample variance on $n(z)$. The result is shown in figure \ref{fig:zdist}. 

\subsubsection{DES Cluster Catalog}
We generate the cluster catalog by running \redmapper{} on the Cardinal gold galaxy catalog. The process is similar to that presented in \citep{cardinal}. In this paper, we bin the \redmapper{} catalog into three tomographic bins based on $z_\lambda$ with bin edges $[0.2, 0.4, 0.55, 0.65]$ to maximize the cross-correlation signals between \redmapper{} and \maglim{}. Within each redshift bin, we further split the samples into four richness bins with bin edges $[20, 30, 40, 60, 500]$. The resulting redshift distribution is shown in the left panel of figure \ref{fig:zdist}.

\subsubsection{Data vectors}
\label{sec:datavec}
In this paper, we consider all possible auto- and cross-correlations between \maglim{} galaxies, \redmapper{} clusters, and shears measured from source galaxies. This includes cosmic shear, galaxy clustering, galaxy--galaxy lensing, cluster--galaxy clustering, cluster clustering, and cluster lensing. The first three constitute the data vector for the \ttt{} analysis \citep{Y6model}. The galaxy clustering, cluster--galaxy clustering, cluster clustering, and cluster lensing constitute the data vector for the \clustercomb{} analysis. To isolate the information, we only consider \maglim{} galaxies that have overlapping redshift ranges as galaxy clusters (i.e., the first three redshift bins) in the \clustercomb{} analysis.

We measure the auto- and cross-correlations of \maglim{} galaxies and \redmapper{} clusters using the Landy-Sazalay estimator \citep{1993ApJ...412...64L}. Specifically, we evaluate \begin{equation}
    \hat{w}(\theta) = \frac{DD-2DR+RR}{RR}, 
\end{equation}
where $DD$ is the number of pairs of tracers (galaxies or galaxy clusters) with angular separation $\theta$, $RR$ is similarly defined for a catalog of points whose positions are randomly distributed within the survey volume (random points), and $DR$ is the number of cross pairs between tracers and random points. The cluster lensing  $\gamma_c (\theta)$ is calculated via the same estimator as implemented in \citep{4x2pt1}, reading as 
\begin{eqnarray}
\label{eq:clusterlensing}
        \hat{\gamma_c}(\theta) 
    &=& \frac{N_r}{N_c}\frac{\sum_{j\in DS(\theta)}e_{Ts,j}}{RS(\theta)} -  \frac{\sum_{j\in RS(\theta) }e_{Ts,j}}{RS(\theta)}, 
\end{eqnarray}
where $DS(\theta)$ is the number of cluster--source galaxy pairs with angular separation $\theta$ and $e_{Ts,j}$ is the tangential shear of the source galaxies in cluster--source galaxy pair, $N_r$ is the number of random points, $N_c$ is the number of galaxy clusters,  and  $RS(\theta)$ is the number of random points--source galaxy pairs with angular separation $\theta$. 

Following equation \ref{temaki}, we further transform the measurement $\gamma_c (\theta)$ to $\Sigma (\theta)$ with a linear transformation. This makes our modeling more resilient to unknown small-scale physics that shapes the matter distribution below the scale of interest.

\begin{figure*}
    \centering
    \includegraphics[width=0.9\textwidth]{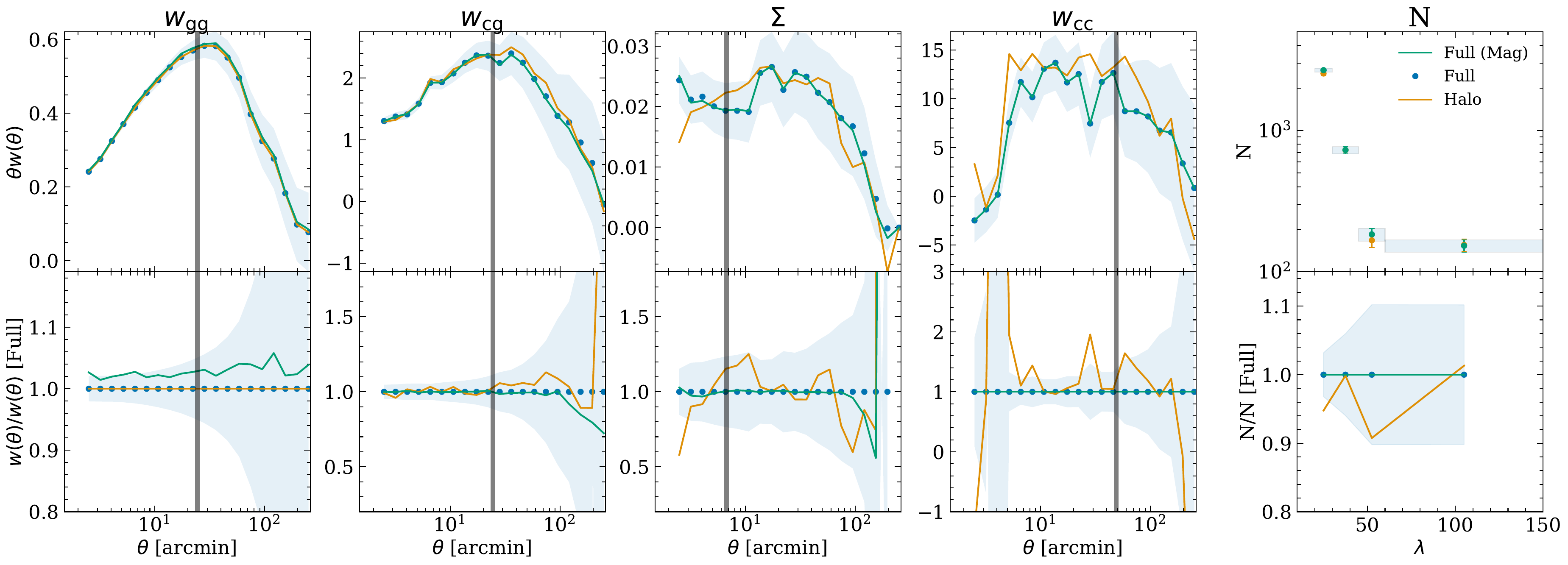}
    \caption{Comparison between the data vector generated from three variants of Cardinal described in section \ref{sec:datavec}: Halo (orange), Full (blue), and Full Mag (green). For clarity, here we show one redshift and one richness bin for each component of the data vector. The errorbars correspond to the $1\sigma$ uncertainty estimated with theoretical calculation detailed in section \ref{sec:model}.}
    \label{fig:datavector}
\end{figure*}

All two-point correlation functions are calculated in 20 logarithmic angular bins between 2.5 to 250 arcmins using \textit{Treecorr} \citep{2004MNRAS.352..338J}. However, due to the limitation of the model, only limited ranges of scales are used for the analysis. Specifically, we only use scales $>8\ h^{-1}{\rm Mpc}$ for galaxy clustering and galaxy--cluster cross correlations, $>16\ h^{-1}{\rm Mpc}$ for cluster clustering, and $>2\ h^{-1}{\rm Mpc}$ for cluster lensing. These scales are defined at the mean redshifts of each tomographic bin, assuming a fiducial cosmology, which is the fiducial value shown in table \ref{tab:params} for simulated likelihood analyses and Cardinal cosmology for Cardinal analyses.

To isolate the different modeling components that shape the two-point correlation functions, we generate the data vector with the following variants of catalogs: 
\begin{enumerate}[label=\Alph*.]
    \item \textbf{Halo:} We use \redmapper{} to count the richness of each dark matter halo in Cardinal. We select \maglim{} and source galaxies using each galaxy's unmagnified photometry. We generate a data vector using the positions of objects that are not deflected due to lensing.  We note that the cluster sample in this run still suffers the projection effect. \redmapper{}'s richness will still include foreground and background galaxy that satisfies the red-sequence criteria.
    \item \textbf{Full:} Similar to variant A, but here we use \redmapper{} to find galaxy clusters. We generate the data vector using the found clusters and their \redmapper{} identified redshift. 
    \item \textbf{Full (Mag):} Similar to variant B, but here, we generate \maglim{} and source galaxies using magnified photometry. The data vector is generated using the positions of objects deflected due to lensing. 
\end{enumerate}
Comparisons of A and B test the validity of \redmapper{} as a cluster finder in the Cardinal universe. Comparisons of B and C test the validity of lensing implementation in Cardinal and the model adopted to mitigate the magnification biases. The measurements of all three variants are shown in figure \ref{fig:datavector}. 

We find that at a given richness, using \redmapper{} as a richness calculation tool (halo run hereafter) leads to slightly fewer clusters than using \redmapper{} as a cluster finder (full run hereafter). This is likely due to miscentering, which leads to a small difference in the richness--mass relations of the two samples. The fewer clusters in the halo run indicate that they are likely to have a larger mass, which is consistent with the fact that the halo run clusters have a slightly larger $w_{\rm cg}$, $\Sigma$, and $w_{\rm cc}$. Despite this small difference in the richness--mass relations, we find that the halo run and full run produce data vectors that are consistent with each other, indicating the excellent performance of \redmapper{} as a cluster finder in Cardinal. 

Finally, we estimate the redshift errors in \maglim{} and source galaxies by comparing mean redshifts estimated using photometric redshift distribution and true redshift distribution. We use the measured differences as the central value of the Gaussian prior for the parameters associated with those redshift uncertainties. We summarize the priors and parameters used for cosmological analyses in table \ref{tab:params}.

\section{Validation of the model}
\begin{table*}[h!]
    \centering
    \footnotesize
    \caption{Parameters and priors considered in this analysis. Flat represents a flat prior in the given range, and $\rm{Gauss}(\sigma)$ denotes a Gaussian prior with width $\sigma$. The means of the Gaussian priors are determined by comparing true redshifts and photometric redshifts of galaxies, thus varying between different versions of simulations. \\}
    \label{tab:params}
   \begin{tabular}{ lccccc}
    \hline 
	Parameter	 & Prior & Fiducial value & Varied in \clustercomb{} & Varied in \allcomb{} \\
	\hline 
	&Cosmology& \\
	$\omegam$ & Flat (0.1,0.9) & 0.3 & \checkmark{} &\checkmark{}\\
	$A_s$ & Flat ($5\times10^{-10}$,$5\times10^{-9}$ ) & 2.19e-9 & \checkmark{} &\checkmark{}\\
	$n_s$ & Flat (0.87, 1.07) & 0.96859& \checkmark{} &\checkmark{}\\
	$\Omega_b$ &  Flat (0.03, 0.07) &0.048& \checkmark{} &\checkmark{}\\
	$h$ &  Flat (0.55, 0.91)&0.69& \checkmark{} &\checkmark{} \\
    $\Omega_{\nu}h^2$ &  Flat (0.0006, 0.00644)& 0.00083& \checkmark{} &\checkmark{}\\
	\hline 
	&Galaxy Bias& \\
	$b_{1,\rm{l}}^1$ &  Flat (0.8, 3.0) &1.42& \checkmark{} &\checkmark{}\\ 
	$b_{1,\rm{l}}^2$ &  Flat (0.8, 3.0) &1.66& \checkmark{} &\checkmark{}\\ 
	$b_{1,\rm{l}}^3$ &  Flat (0.8, 3.0) &1.7& \checkmark{} &\checkmark{}\\ 
    $b_{1,\rm{l}}^4$ &  Flat (0.8, 3.0) &1.62& - &\checkmark{}\\ 
	$b_{1,\rm{l}}^5$ &  Flat (0.8, 3.0) &1.78& - &\checkmark{}\\ 
	$b_{1,\rm{l}}^6$ &  Flat (0.8, 3.0) &1.75&-  &\checkmark{}\\ 
	\hline
 &Intrinsic alignment& \\
 $a_1$ & Flat (-5.0, 5.0)& 0.0& \checkmark{} &\checkmark{} \\
  $\eta_1$ & $\rm{Gauss}(\mu=0, \sigma=3)$  & 0.0& \checkmark{} &\checkmark{} \\
  $a_2$ & Flat (-5.0, 5.0)& -1.36& - &\checkmark{} \\
  $\eta_2$ & $\rm{Gauss}(\mu=0, \sigma=3)$ & 0.0& -&\checkmark{}  \\
 \hline
	&\maglim{} photo-$z$&\\
	$\Delta_{z,\rm{l}}^1$ & $\rm{Gauss}(\mu=0.005, \sigma=0.007)$&0.005& \checkmark{} &\checkmark{} \\
	$\Delta_{z,\rm{l}}^2$ & $\rm{Gauss}(\mu=0.003, \sigma=0.011)$&0.003 & \checkmark{} &\checkmark{}\\
	$\Delta_{z,\rm{l}}^3$ & $\rm{Gauss}(\mu=0.001, \sigma=0.006)$&0.001 & \checkmark{} &\checkmark{}\\
	$\Delta_{z,\rm{l}}^4$ & $\rm{Gauss}(\mu=-0.002, \sigma=0.006)$&-0.002 & -&\checkmark{}\\
	$\Delta_{z,\rm{l}}^5$ & $\rm{Gauss}(\mu=0.001, \sigma=0.01)$&0.001 & - &\checkmark{}\\
	$\Delta_{z,\rm{l}}^6$ & $\rm{Gauss}(\mu=0.008, \sigma=0.01)$&0.008& - &\checkmark{} \\
	\hline
    &\maglim{} magnification&\\
	$C_{\rm{l}}^1$ & fixed & -1.57& - &-\\
	$C_{\rm{l}}^2$ &fixed& -1.70& - &-\\
	$C_{\rm{l}}^3$ &fixed&-0.25& - &- \\
        $C_{\rm{l}}^4$ & fixed & 1.50& - &-\\
	$C_{\rm{l}}^5$ &fixed& 2.22& - &-\\
	$C_{\rm{l}}^6$ &fixed&2.80& - &- \\
    	\hline
    &point mass marginalization&\\
	$B^i$ & Flat (-1.0, 1.0) & 0.0& - &\checkmark{}\\
	\hline
	&Source galaxy photo-$z$& \\
	$\Delta_{z,\rm{s}}^1$& $\rm{Gauss}(\mu=0.034, \sigma=0.018)$ &0.034 & \checkmark{} &\checkmark{}\\
	$\Delta_{z,\rm{s}}^2$ &$\rm{Gauss}(\mu=0.028, \sigma=0.013)$ &0.028& \checkmark{} &\checkmark{}\\
	$\Delta_{z,\rm{s}}^3$& $\rm{Gauss}(\mu=0.011,\sigma=0.006)$ & 0.011& \checkmark{} &\checkmark{}\\
	$\Delta_{z,\rm{s}}^4$& $\rm{Gauss}(\mu=-0.010, \sigma=0.013)$ & -0.010& \checkmark{} &\checkmark{}\\
		\hline
 &Shear Calibration& \\
         $m_1$ & $\rm{Gauss}(\mu=0, \sigma=0.008)$ &0 & \checkmark{} &\checkmark{} \\
$m_2$ & $\rm{Gauss}(\mu=0, \sigma=0.013)$ &0& \checkmark{} &\checkmark{}\\
$m_3$ & $\rm{Gauss}(\mu=0, \sigma=0.009)$&0& \checkmark{} &\checkmark{} \\
$m_4$ & $\rm{Gauss}(\mu=0, \sigma=0.012)$ &0& \checkmark{} &\checkmark{}\\
 \hline
    &\redmapper{} richness--mass relation \\
    $\rm{ln}\lambda_{0}$ &  Flat (2.0,5.0) & 4.26& \checkmark{} &\checkmark{}\\
    $A_{\rm{ln}\lambda}$ &  Flat (0.1,1.5) &0.943& \checkmark{} &\checkmark{}\\
    $B_{\rm{ln}\lambda}$ & Flat(-5.0, 5.0) &0.207& \checkmark{} &\checkmark{}\\
    $\sigma_{\rm{intrinsic}}$ &  Flat(0.1, 1.0) &0.15& \checkmark{} &\checkmark{}\\
    \hline 
    &\redmapper{} selection effect \\
    $b_{s1}$ &  Flat (1.0,2.0) &1.1& \checkmark{} &\checkmark{}\\
    $b_{s2}$ & Flat(-1.0,1.0) &0.2& \checkmark{} &\checkmark{}\\
    $r_0$ & Flat(10, 60) & 30& \checkmark{} &\checkmark{}\\
        &\redmapper{} magnification&\\
	$C_{\rm{c}_A}^i$ &fixed& -2&-&-\\
	\hline
   \hline \vspace{-3mm}\\   
    \end{tabular}
\end{table*}

In this section, we present the main results of the paper --- we show that the model used in this paper is robust for the \clustercomb{} analysis with DES Y6 data, both in a simulated likelihood analysis (Section~\ref{sec:simulated}) and an end-to-end analysis carried out with the Cardinal mock catalogs (Section~\ref{sec:mocktest}). We further show that possible shared systematics between \clustercomb{} and \ttt{} will not bias the \allcomb{}  using simulated likelihood analyses (Section~\ref{sec:simulated}).

\subsection{Simulated likelihood analysis}
\label{sec:simulated}

The model presented in Section~\ref{sec:model} is the fiducial model we will use in the forthcoming DES Y6 \clustercomb{} analysis. Before we can apply the model to data, it is important to check that 1) with a simulated noiseless data vector, we recover the input cosmology, and 2) our results are robust to the most dominant systematic effects in the model.

We quantify the impact of known unmodeled systematics on our cosmological constraints. Specifically, we consider the following systematics:
\begin{enumerate}
    \item The functional form of the richness--mass relation for the cluster sample (\textbf{MOR}). 
    \item Uncertainties in halo mass function and halo bias (\textbf{Emu}).
    \item Impacts of baryonic effect on the halo mass function, halo bias, and matter power spectra (\textbf{Hydro}). 
    \item Non-linear galaxy overdensity--matter overdensity relation, or the nonlinear galaxy bias (\textbf{NL Bias}).
    \item Non-linear matter clustering (\textbf{NL matter clustering}).
    \item Selection effects in the cluster sample (\textbf{Selection Effect}). 
    \item Systematics in the one-halo regime of cluster lensing (\textbf{One-halo lensing}).
\end{enumerate}
We will use the shorthand in the parentheses to refer to each effect below.

We employ the simulated likelihood analysis technique, where we analyze synthetic data vectors assuming the fiducial model presented in Section~\ref{sec:model} and a Gaussian likelihood with a covariance matrix presented in Section~\ref{sec:cov}. The synthetic data vector is generated with the fiducial model assuming fixed cosmological and nuisance parameters (shown in Table~\ref{tab:params}) and Cardinal redshift distributions shown in Figure~\ref{fig:zdist}. The covariance matrix is generated with DES-Y6-like shape noise, which is $\sim 10\%$ lower than the unblinded shape noise in DES-Y6 \cite{Y6Shear}, and DES-Y6-like number density of \maglim{} galaxies, which is $\sim 10\%$ higher than the unblinded number density in DES-Y6 \cite{Y6Lens}. We note that this mismatch makes the simulated likelihood analysis more conservative.  

To test the impact of systematics on the parameter constraints, we compare the parameter estimation from theoretically contaminated synthetic data vectors and from uncontaminated synthetic data vectors. We compare with the result from the uncontaminated synthetic data instead of the true cosmological parameters because the marginalized parameter posteriors may appear biased from the true parameter values due to degeneracy in high dimensional parameter spaces, known as the prior volume effect \citep{y3-generalmethods}. We require the shifts in $\omegam$ and $\sigma_8$ plane to be less than $0.3$ times the expected statistical uncertainties for each systematic to ensure systematic uncertainties well below the statistical uncertainties. Below, we detail the generation of theoretically contaminated synthetic data vectors.

\subsubsection{Functional form of the richness--mass relation}
The richness--mass relation depends on the properties of galaxies in clusters and the performance of the cluster finder, which is sensitive to the contrast of galaxy properties in clusters and the field. Thus, the richness--mass relation is sensitive to the statistical relations between luminosities, colors, and positions of galaxies and their host halo masses across a wide range of halo mass and redshifts, making it hard to make a theoretical prediction. Here, we follow the approach adopted in DES-Y1 \citep{4x2pt1}, where we validate the sensitivity of our analysis to the different choices of the richness--mass relation. Specifically, we generate synthetic data assuming the richness--mass relation in \citep{DES_cluster_cosmology}, where the authors assume two contributions to the richness--mass relation of \redmapper{} clusters. The first contribution comes from the intrinsic galaxy abundance--halo mass relation, which is parametrized as a Poisson distribution following the usual Halo Occupation Distribution prescription. The second contribution comes from projected halos whose galaxies are miscounted due to photometric redshift uncertainties. This second contribution is calibrated using gravity-only simulations, where the photometric redshift uncertainties are estimated from DES-Y1 data. We note that the photo-$z$ uncertainty in DES-Y6 data is expected to be better than in DES-Y1 data, making this calculation a conservative assumption.

\subsubsection{Uncertainties in halo mass function and halo bias}
\label{ssec:emu_test}
The assumed the Tinker halo mass function and halo bias--halo mass relation are known to have systematic uncertainties of $5-10\%$ and $10-20\%$ respectively \citep{HMFemulator, biasemu}.  To estimate their impact on cosmological constraints, we generate the synthetic data vector using the {\sc Aemulus} emulators based on gravity-only simulations \citep{HMFemulator, biasemu}, which reduces these uncertainties to $1\%$ for the halo mass function and $3\%$ for the halo bias--halo mass relation.  We then analyze this synthetic data vector with the Tinker halo mass function and halo bias--halo mass relation model. 

\subsubsection{Impacts of baryonic effect on the halo mass function, the halo bias, and the matter power spectra}
\label{ssec:Pmm_test}
The halo mass function, the halo bias--halo mass relations, and the matter power spectra can be affected by baryonic physics. Since the impact of the baryonic effect is expected to affect the one-halo regime \citep{godmaxpaper}, where the modeling uncertainties are explicitly tested for galaxy cluster analyses (\ref{subusubsec:onehalo}), we do not test the impact of baryonic effects on matter power spectra for cluster analyses. On the other hand, we test the impact of baryonic effects on the modeling of the halo mass function and halo bias explicitly. To test this, we generate synthetic data using the fitting functions for the halo mass function and halo bias provided in \citep{2021MNRAS.500.2316C}, which is based on the Magneticum simulation \citep{2013MNRAS.428.1395B}. For the specific parts of the model that are affected, please see equations~\ref{eq:bias} and ~\ref{eq:hm}.

Baryonic effects on matter power spectra will affect clusters and \ttt{} simultaneously. While we do not expect these effects will impact cluster-only analyses, they might affect joint constraints of clusters and \ttt{}. To validate this, we generate synthetic data using HMcode2020 with $T_{\rm AGN}=8.0$, which corresponds to a strong baryonic effect similar to the BAHAMAS 8.0 simulation. We note that tests of the strong baryonic effect using the BAHAMAS 8.0 simulation and no baryonic effect for \ttt{} are presented in \citep{Y6model}.

\subsubsection{Nonlinear galaxy overdensity--matter overdensity relation}
\label{ssec:NL_test}
We model the contribution of non-linear galaxy bias to galaxy clustering, galaxy--cluster cross-correlations, and cluster-clustering using an effective 1-loop power spectrum model with non-linear bias parameters: $b_2$ (local quadratic bias), $b_s^2$ (tidal quadratic bias), and $b_{\rm 3nl}$ (third order non-local bias). The general form of this contribution can be written as 
\begin{eqnarray}
\left.\begin{aligned}
     &P_{\rm{1loop}, A,B}(k,z) =  \\
     & \frac{1}{2}
     (b_{1A}b_{2B}+b_{1B}b_{2A})P_{b_1b_2}(k)+\frac{1}{4}b_{2A}b_{2B}P_{b^2b^2}(k)\nonumber \\
     &+\frac{1}{2}(b_{1A}b_{s^2B}+b_{1B}b_{s^2A})P_{b_1s^2}(k)\\&
     +\frac{1}{4}(b_{2A}b_{s^2B}+b_{2B}b_{s^2_A})P_{b_2s^2}(k)\nonumber\\
     &+\frac{1}{2}(b_{1A}b_{3\rm nlB}+b_{1B}b_{\rm 3nlA})P_{b_1b_3}(k)+\frac{1}{4}(b_{s^2A}b_{s^2B})P_{s^2s^2}(k), \\
 \end{aligned}\right.
\end{eqnarray}
where $A,B$ denotes the tracers (e.g., clusters, galaxies, or matters), and $P_{b_1b_2}$, $P_{b_2b_2}$, $P_{b_1s^2}$, $P_{b_2s^2}$, $P_{b_1b_3}$, and $P_{s^2s^2}$ are integrals of power spectra evaluated using FAST-PT code \citep{2016JCAP...09..015M}. Note that if the tracer is matter, we have $b_1=1, b_2=0, b_{s^2}=0, b_{3\rm nl}=0$. As for galaxies and clusters, following the analysis choice of \citep{DESmethod}, we model $b_{s^2}$ and $b_{3\rm nl}$ using their co-evolution value 
\begin{eqnarray}
    b_{s^2} &=& \frac{-4}{7}(b_1-1) \nonumber \\
    b_{3\rm nl} &=& b_1-1.
\end{eqnarray}
We adopt the $b_2$ of \maglim{} galaxies determined in the \ttt{} analysis \citep{Y3kp}. For clusters, we found that adopting the $b_2$--$b_1$ relation measured in separate universe N-body simulations \citep{2016JCAP...02..018L} overpredicts the measured projected cluster--clustering signal (See Appendix~\ref{app:NL} for details). Because this overprediction could lead to an overly conservative conclusion, we adopt a different approach from DES-Y1 \citep{4x2pt1}. Specifically, as detailed in Appendix~\ref{app:NL}, we adopt the $b_2$ value for clusters obtained from fitting the non-linear bias model to the halo clustering signal measured in the N-body simulation.  

Similarly, we add the 1-loop contribution to galaxy--galaxy lensing in the \allcomb{} and \allcomb{}+BAO+SN data vector. Specifically, we add the following terms to the galaxy--matter power spectra, 
\begin{eqnarray}
\left.\begin{aligned}
     &P_{\rm{1loop}, g, m}(k,z) =  \\
     & \frac{1}{2}
     b_{2g}P_{b_1b_2}(k)+\frac{1}{2}b_{s^2g}P_{b_1s^2}(k)+\frac{1}{2}b_{3\rm nlg}P_{b_1b_3}(k).\\
 \end{aligned}\right.
\end{eqnarray}

We further test the impact of non-linear galaxy overdensity--matter overdensity relation on cluster lening.  Since our modeling already incorporates the one-halo term, directly adding non-linear bias contributions would lead to double counting. To account for this effect consistently, we adopt predictions from N-body simulations. Specifically, we use the Dark Emulator \citep{2019ApJ...884...29N} to generate non-linear matter-halo power spectra, which we then integrate into our theoretical framework to construct the cluster lensing data vector.

\subsubsection{Non-linear matter clustering}
\label{ssec:NL_test_matter}
We model the non-linear matter clustering using the Euclid Emulator2 \citep{EuclidEmulator2}. Because Euclid Emulator2 only predicts the nonlinear matter clustering for a gravity-only universe, we analyze this data vector with the gravity-only HMcode2020 instead.

\subsubsection{Selection effect model}
We adopt the selection effect model presented in \cite{Sunayama2023, Park2023}\footnote{During the analysis, we found typos in \cite{Sunayama2023, Park2023}. The $\Pi(R)$ in those papers should be $1+\Pi(R)$.}. Specifically, we adopt the best-fit value of HSC-Y1 lensing and SDSS \redmapper{} cluster abundance analyses \cite{Sunayama2023}. We apply the following conversion to the cluster--galaxy cross-correlation ($w_{cg}(\theta)$), localized cluster lensing ($\Sigma(\theta)$) and cluster clustering ($w_{cc}(\theta)$)
\begin{eqnarray}
    w_{cg}(\theta) &=& \Pi(\theta) w_{cg}^{org}(\theta),\\
    w_{cc}(\theta) &=& \Pi^2(\theta) w_{cc}^{org}(\theta),\\
    \Sigma(\theta) &=& \Pi(\theta) \Sigma^{org}(\theta), 
\end{eqnarray} 
where $\Pi(\theta)$ is defined as 
\begin{eqnarray}
    \Pi(\theta)= 
    \begin{cases}
    1+\Pi_0 (R(\theta)/R_0) & \text{if } \theta \leq R_0,\\
    1+\Pi_0(\theta)+c\ln(\theta/R_0),               & \text{if } \theta>R_0
\end{cases}
\end{eqnarray}
where, $R(\theta)$ is the projected comoving separation at the mean redshift of the clusters, $R_0$ is converted from the best-fit value in \cite{Sunayama2023} at the mean redshift of each tomographic bin, assuming the fiducial cosmology in the simulated likelihood analyses.

\subsubsection{Possible systematics in the one-halo regime of cluster lensing}
\label{subusubsec:onehalo}
In this test, we artificially make the one-halo term cluster lensing $50\%$ smaller in the contaminated data vector to test the sensitivity of our analysis to the small-scale cluster lensing modeling. This test encompasses a number of possible small-scale systematics: baryonic impact on cluster lensing, mis-centering, and scale-dependent projection effects.

\subsubsection{Summary of simulated likelihood analyses}
\begin{figure*}
    \centering
    \includegraphics[width=1.0\textwidth]{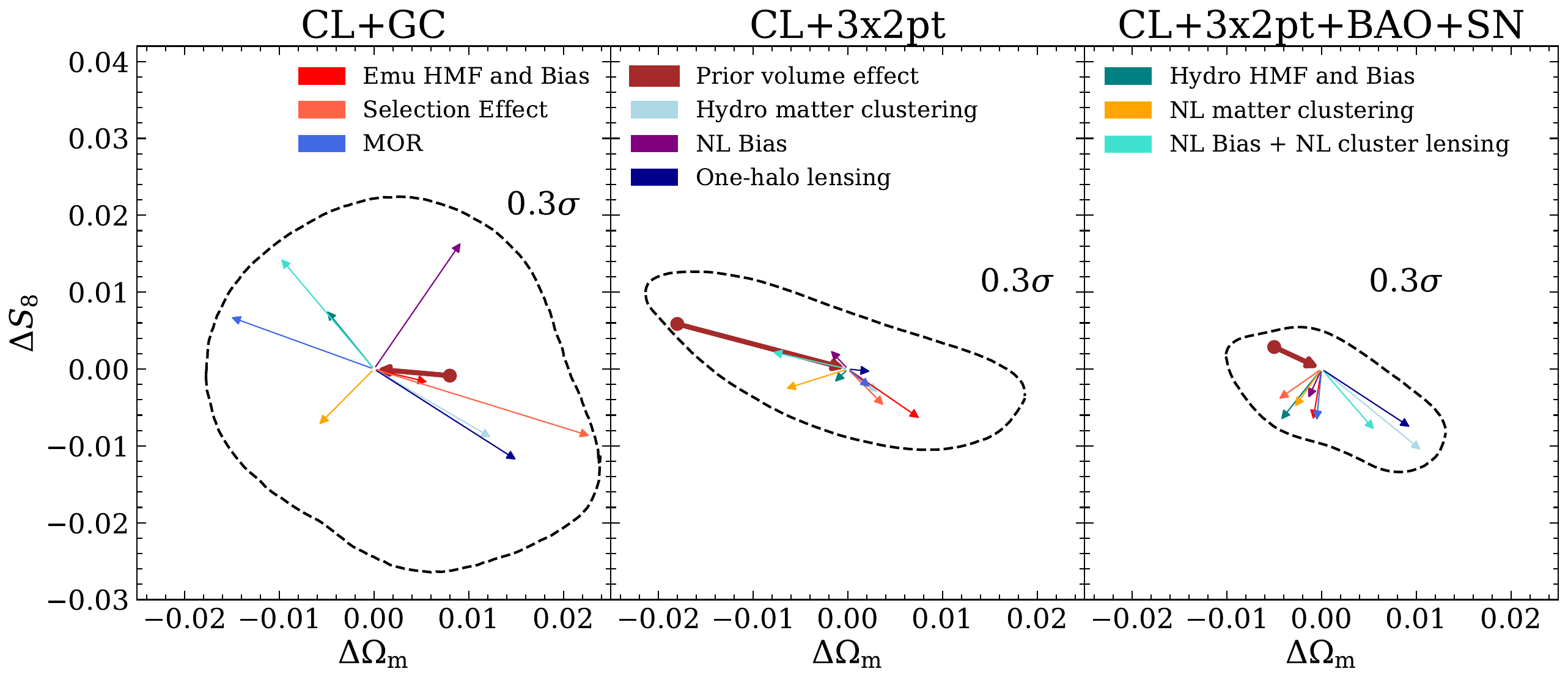}
    \caption{ Parameter biases for \clustercomb{} ({\em left}), \allcomb{} ({\em middle}), and  \allcomb{}+BAO+SN ({\rm right}) analyses due to different systematics: arrows show the difference between marginalized constraints using the baseline and contaminated data vector. The dashed line shows $0.3\sigma$ confidence intervals of the baseline analyses. Different colors correspond to different systematics detailed in section \ref{sec:simulated}. The length of each arrow represents the magnitude of the bias introduced by each contamination. Except for the prior volume effect, the arrow directions indicate the bias relative to the analysis of the true data vector. The arrow for the prior volume effect, however, shows the bias of the analysis on the true data vector relative to the true cosmology used to generate it. 
    }
    \label{fig:simulated}
\end{figure*}
Figure \ref{fig:simulated} summarizes the test result for  \clustercomb{}, \allcomb{}, \allcomb{}+BAO+SN. For each test, we show both the amplitude and direction of the bias on $S_8$ and $\omegam$ for each systematic. First, for the cluster analysis, we find that systematics shift the cosmological parameters in different directions. Some of the directions are interpretable.  For example, the non-linear galaxy bias boosts the power spectra, which can be compensated with higher $S_8$. While others are hard to understand intuitively due to the complicated nature of the data vector, we find that none of the above systematics can bias the cosmological constraint by more than $0.3$ of the statistical uncertainties.  Interestingly, we find an impact on cosmological constraints due to the baryonic effect on halo mass function and halo bias as described by \cite{2021MNRAS.500.2316C}. If the baryonic effect simply changes the mass of the halos but preserves the halo number density--halo bias relation, we would observe no impact on cosmological constraints. While tantalizing, a detailed investigation of where and how the model breaks the halo number density--halo bias relation is beyond the scope of this work.

We extend all systematic contamination tests to the combined analyses of clusters and \ttt{}, as well as clusters, \ttt{}, and BAO+SN. Naively, one might assume that it is only necessary to test systematics that affects all data vectors. However, as demonstrated in Appendix \ref{app:biasmultiprobe}, incorporating additional probes that are individually free of systematics and uncorrelated with the primary probe can still amplify the impact of systematics in the primary probe on the total error budget due to degeneracy breaking. To account for this, we test all identified systematics across the clusters and \ttt{} joint analysis, as well as the combined analysis of clusters, \ttt{}, and BAO+SN. The result is summarized in the middle and right panel of figure~\ref{fig:simulated}. As expected, the baryonic effect on matter clustering tends to bias $S_8$ low because baryonic feedback pushes gas outside of halos. The bias due to nonlinear galaxy bias shows relatively small changes in $S_8$ compared to \clustercomb{}. This is likely because cosmic shear, which does not depend on the galaxy bias modeling, dominates the constraining power on $S_8$. Further, figure~\ref{fig:simulated} shows that none of the tested systematic will bias constraints by more than $0.3$ of the statistical uncertainties. We thus conclude that our model is sufficiently flexible to enable us to derive robust cosmological constraints at the precision achievable by DES Y6-like surveys.

\subsection{Cardinal mock test}
\label{sec:mocktest}
\begin{figure}
    \centering
    \includegraphics[width=0.5\textwidth]{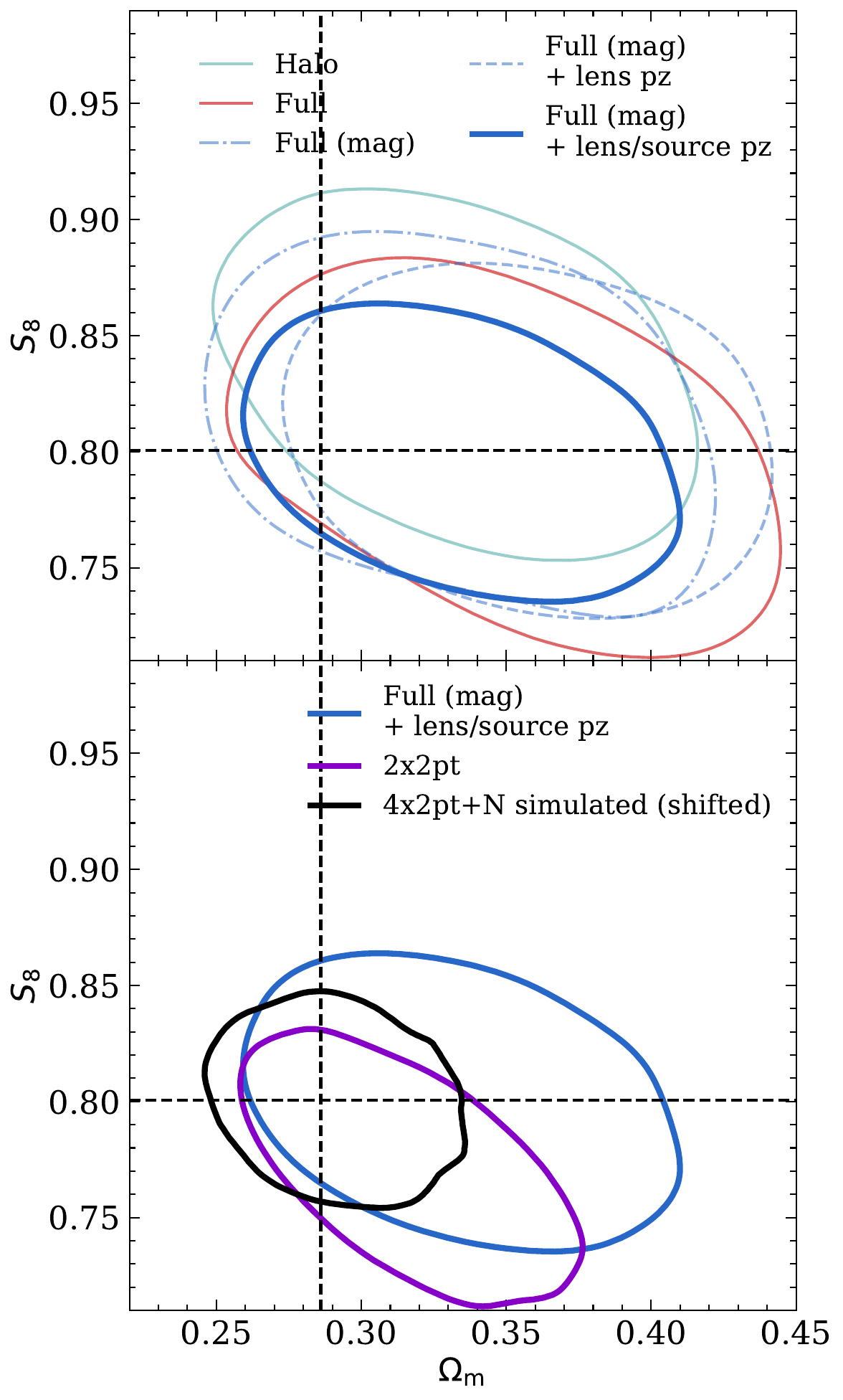}
    \caption{ $\Omega_{\rm m}$ and $S_8$ constriants in \clustercomb\ analyses in different Cardinal runs. The contours show the $68\%$ confidence levels, corresponding to the $1\sigma$ statistical error. Analyses are performed on three Cardinal catalogs. Green contour is performed on the Halo run. Red contour is performed on the Full run. Blue contours are performed on the Full (Mag) run. The blue contour with a dot-dashed line corresponds to an analysis using true redshift of lens and source galaxies, the blue contour with a dashed line considers lens photometric redshift uncertainties, and the blue contour with a solid line considers lens and source photometric redshift uncertainties. As a comparison, in the \textit{Lower} panel, the purple contour shows constraints from galaxy--galaxy lensing and galaxy clustering analyses ($2\times 2\rm{pt}$). The black contour shows the expected constraining power from DES-Y6 data. The dashed-black line indicates the true cosmology used to generate the Cardinal simulation.  }
    \label{fig:cardinal}
\end{figure}
Finally, we perform end-to-end cosmology recovery tests using the Cardinal simulation. While several systematics have been tested with simulated likelihood analyses, the tests on Cardinal provide an overall estimation of the combined impact of some of the systematics. Further, the systematics in the selection-effect model and the mass--observable relation lead to the largest shift in cosmological parameters for cluster analyses. These two systematics are also expected to be correlated because projected line-of-sight structures can change the richness in a given halo and the associated two-point correlation functions. It is, therefore, important to test them simultaneously. Here, we use the Cardinal simulation as an alternative test of those models. Finally, because we can run \redmapper{} on Cardinal in the same way as it is run on data, we can use it to test \redmapper{}'s cluster finding performance. 

To carry out this test, we employ the \clustercomb{} data vector measured in three Cardinal catalogs described in section~\ref{sec:datavec}: Halo, Full, and Full (Mag). We then carry out our cosmological analysis program described in section~\ref{sec:model} on the simulated universe. We further make several simplifications. First, the shift in the mean of photometric redshifts in Cardinal is measured by comparing photometric redshifts and true redshifts of galaxies instead of estimating from empirical methods. Second, the covariance matrix is calculated using the true cosmology instead of estimating it through an iterative process. Third, we estimate the magnification bias of galaxies in Cardinal using the $\kappa$ measured in Cardinal directly. While Cardinal will not shed light on methods of quantifying redshift biases, iterative processes in covariance matrix generations, and estimation of magnification biases, we note that each of these methods has been well tested \citep{y6ClusteringRedshift, y3-covariances, Y6Balrog}.

We focus on $S_8$ and $\omegam$ constraints in the tests on Cardinal. The result is shown in Figure~\ref{fig:cardinal}, where we show  $1\sigma$ contours of cosmological analyses of all three Cardinal catalogs described in Section~\ref{sec:datavec}: Halo, Full, Full (Mag). Overall, we find good agreement between different variations, and all variations recover the true cosmological parameters within $1\sigma$ confidence interval. The agreement between the Full (red) and Halo (green) runs tests the validity of \redmapper{} as a cluster finder. The agreement between the Full and Full (Mag) runs suggests the validity of our magnification model for galaxies and clusters. We further test the impact of photometric redshift uncertainties and mitigation methods. Specifically, we make two additional runs using photometric redshifts of lens galaxies (blue dashed) and source galaxies (blue solid). The agreement between different photo-$z$ runs suggests our models of photometric uncertainties for lens and source galaxies are sufficient to absorb the potential biases in $n(z)$ that impact \clustercomb{} cosmological constraints. Note that the constraining power is worse than those in the simulated analyses due to the $50\%$ differences in cluster abundance in Cardinal and the data caused by a mismatch of \redmapper{} versions. Because of this caveat and because there is only one Cardinal simulation, the noise in Cardinal can be so large that it hinders the discovery of possible systematics. We, therefore, compare the \clustercomb{} constraints to the joint analyses of galaxy clustering and galaxy-galaxy lensing constraints ($2\times2\rm{pt}$) using galaxies in the same redshift ranges as the cluster analyses on the same realization. Comparison of the differences between \clustercomb{} and $2\times 2 \rm{pt}$ reduces the impact of cosmic variance. We find that the deviation between \clustercomb{} and $2\times2\rm{pt}$ on $S_8$ and $\omegam$ is $0.447\sigma$, which is similar to what we find in DES-Y1 \citep{4x2pt1}. %

Overall, our results on Cardinal show no detection of model misspecification for \clustercomb{} in the Cardinal universe. While the simulated likelihood tests ensure that each of the tested systematics will not be detectable in our Cardinal tests, those tests do not guarantee that the combination of all combined systematics will not be detectable in Cardinal. Thus, Cardinal provides an additional test for the overall systematics of our modeling plan. This overall systematic budget includes the combined systematics in the modeling of \clustercomb{}, the performance of \redmapper{} cluster finding algorithm, the performance of photometric redshift estimations, the method to mitigate photometric redshift biases, and the modeling of the impact of magnification.

\section{Conclusion}
\label{sec:conc}
In this paper, we outline an analysis program for joint analyses of cluster abundances, galaxy clustering, gravitational lensing, BAO, and type-Ia supernovae. We employ the simulated likelihood technique and catalog-level DES simulations to quantify the modeling accuracy of cluster-related data vectors (\clustercomb{}). We further validate the modeling accuracy for the joint analyses of clusters and other cosmological probes in DES year 6 data. We note that the non-cluster-related pipeline has been validated in companion work \citep{Y6model, Y5SN, Y6BAO}. While it might seem redundant that some of the tested systematics are performed for \ttt{} \citep{Y6model}, we stress that it is crucial to evaluate their impact on the full combined data vector. This is because degeneracy breaking can amplify the contribution of systematics to the total error budget, as demonstrated in Appendix \ref{app:biasmultiprobe}.

In section \ref{sec:model}, we begin by giving a general description of our likelihood model. We then detail the assumptions in the model we employ in data analyses, which we explicitly validate in section \ref{sec:simulated}. 

In section \ref{sec:mockuniverse}, we detail the process of constructing \clustercomb{} data vector using the Cardinal simulation \citep{cardinal}. We begin by constructing samples that are necessary for \clustercomb{} analyses, including the source galaxy catalog, the lens galaxy catalog, and the galaxy cluster catalog. The lens galaxy catalog (\maglim{}) is generated based on photometric redshift (DNF), which is generated by running the DNF algorithm on Cardinal in a similar way as data. The sensitivity to survey property fluctuation is mitigated by weights generated with the ENET method. The source galaxy catalog is generated with a customized photometry that mimics the data's photometric distribution. We further supplement the source galaxy catalog with redshift binning and photometric redshift estimation using the SOMPZ algorithm. Finally, we generate the cluster catalog by running the \redmapper{} cluster finder. The data vector is then generated with the measurement pipeline used for data analyses on those three catalogs.

In section \ref{sec:simulated}, we detail the calculation of potential modeling systematics for \clustercomb{}, \allcomb{}, and \allcomb{}+BAO+SN. We show that none of those will bias the cosmological constraints by more than $0.3$ of the expected statistical uncertainties. In section \ref{sec:mocktest}, we further use the Cardinal simulations to validate additional modeling uncertainties. While we focus on \clustercomb{} in the Cardinal validation in this paper, we note that the \ttt{}'s validation in Cardinal is presented in \citep{Y6model}. This is facilitated by three variants of Cardinal simulations, which are used to explicitly test cluster finding, magnification biases, and photometric redshift error mitigations. Given all tests done in this paper, we report no detection of possible systematics given the DES Y6 accuracy. As such, our pipeline is sufficiently accurate for performing a $\Lambda$CDM cosmological analyses of the DES Y6 data in the combinations: \clustercomb{}, \allcomb{}, and \allcomb{}+BAO+SN. These analyses will be presented in forthcoming papers.

Looking forward, the next generation of survey data from Euclid, LSST, and Roman is poised to become more constraining than DES Y6 in the next 3--5 years. The study in this paper highlights that more constraining data will push the requirement on the modeling pipeline to the next level.
More constraining surveys will require even more sophisticated and complex models to meet the required modeling accuracy. In addition, the accuracy of these models can only be guaranteed with yet more sophisticated, high-fidelity, large-volume simulations. While improving the fidelity of the modeling is essential, it is equally important to develop high-quality simulations that have sufficient volume to characterize the systematic effects in these future datasets.  This will be a major computational challenge in terms of both the mock catalog generation and the cycles needed to run survey-specific end-to-end pipelines (similar to the Cardinal test done in this paper) multiple times.
To alleviate this hurdle and to avoid delays coming from constantly evolving pipelines, we encourage designing analysis pipelines with a focus on automation and efficiency, enabling them to be rerun multiple times seamlessly.

\section{Acknowledgements.}

{\bf Author Contributions:}
 All authors contributed to this paper and/or carried out infrastructure work that made this analysis possible. Chun-Hao To (CT) performed the overall analysis in this paper and most of the manuscript preparation. Elisabeth Krause (EK), Eduardo Rozo (ER), Heidi Wu (HW), Risa Wechsler (RW), David Weinberg (DW), and CT formed the core discussion group. EK, Chihway Chang (CC), and CT contributed to the manuscript preparation. Sebastian Bocquet and Jonathan Blazek were DES internal reviewers. The remaining authors have made contributions to this paper that include, but are not limited to, the construction of DECam and other aspects of collecting the data; data processing and calibration; developing broadly used methods, codes, and simulations; running the pipelines and validation tests; and promoting the science analysis.

CT is supported by the Eric and Wendy Schmidt AI in Science Postdoctoral Fellowship, a Schmidt Futures program. EK is supported in part by Department of Energy grant DE-SC0020247 and the David and Lucile Packard Foundation. ER was funded by DOE grant DE-SC0009913.  HW was supported by DOE grant DE-SC0021916.
Development of the Cardinal simulations and its predecessors was primarily funded by the U.S. Department of Energy under contract number DE-AC02-76SF00515 to SLAC National Accelerator Laboratory and by Stanford University.
CT would like to thank Stanford University, the Stanford Research Computing Center, the Ohio Supercomputer Center,  and the University of Chicago’s Research Computing Center for providing the computational resources and support that contributed to these research results. 

Funding for the DES Projects has been provided by the U.S. Department of Energy, the U.S. National Science Foundation, the Ministry of Science and Education of Spain, 
the Science and Technology Facilities Council of the United Kingdom, the Higher Education Funding Council for England, the National Center for Supercomputing 
Applications at the University of Illinois at Urbana-Champaign, the Kavli Institute of Cosmological Physics at the University of Chicago, 
the Center for Cosmology and Astro-Particle Physics at the Ohio State University,
the Mitchell Institute for Fundamental Physics and Astronomy at Texas A\&M University, Financiadora de Estudos e Projetos, 
Funda{\c c}{\~a}o Carlos Chagas Filho de Amparo {\`a} Pesquisa do Estado do Rio de Janeiro, Conselho Nacional de Desenvolvimento Cient{\'i}fico e Tecnol{\'o}gico and 
the Minist{\'e}rio da Ci{\^e}ncia, Tecnologia e Inova{\c c}{\~a}o, the Deutsche Forschungsgemeinschaft and the Collaborating Institutions in the Dark Energy Survey. 

The Collaborating Institutions are Argonne National Laboratory, the University of California at Santa Cruz, the University of Cambridge, Centro de Investigaciones Energ{\'e}ticas, 
Medioambientales y Tecnol{\'o}gicas-Madrid, the University of Chicago, University College London, the DES-Brazil Consortium, the University of Edinburgh, 
the Eidgen{\"o}ssische Technische Hochschule (ETH) Z{\"u}rich, 
Fermi National Accelerator Laboratory, the University of Illinois at Urbana-Champaign, the Institut de Ci{\`e}ncies de l'Espai (IEEC/CSIC), 
the Institut de F{\'i}sica d'Altes Energies, Lawrence Berkeley National Laboratory, the Ludwig-Maximilians Universit{\"a}t M{\"u}nchen and the associated Excellence Cluster Universe, 
the University of Michigan, NSF NOIRLab, the University of Nottingham, The Ohio State University, the University of Pennsylvania, the University of Portsmouth, 
SLAC National Accelerator Laboratory, Stanford University, the University of Sussex, Texas A\&M University, and the OzDES Membership Consortium.

Based in part on observations at NSF Cerro Tololo Inter-American Observatory at NSF NOIRLab (NOIRLab Prop. ID 2012B-0001; PI: J. Frieman), which is managed by the Association of Universities for Research in Astronomy (AURA) under a cooperative agreement with the National Science Foundation.

The DES data management system is supported by the National Science Foundation under Grant Numbers AST-1138766 and AST-1536171.
The DES participants from Spanish institutions are partially supported by MICINN under grants PID2021-123012, PID2021-128989 PID2022-141079, SEV-2016-0588, CEX2020-001058-M and CEX2020-001007-S, some of which include ERDF funds from the European Union. IFAE is partially funded by the CERCA program of the Generalitat de Catalunya.

We  acknowledge support from the Brazilian Instituto Nacional de Ci\^encia
e Tecnologia (INCT) do e-Universo (CNPq grant 465376/2014-2).

This document was prepared by the DES Collaboration using the resources of the Fermi National Accelerator Laboratory (Fermilab), a U.S. Department of Energy, Office of Science, Office of High Energy Physics HEP User Facility. Fermilab is managed by Fermi Forward Discovery Group, LLC, acting under Contract No. 89243024CSC000002.
\bibliographystyle{apsrev}
\bibliography{sample}
\clearpage
\onecolumngrid
\appendix
\section{Overall magnitude distribution comparison}
\begin{figure*}
    \centering
    \includegraphics[width=0.9\textwidth]{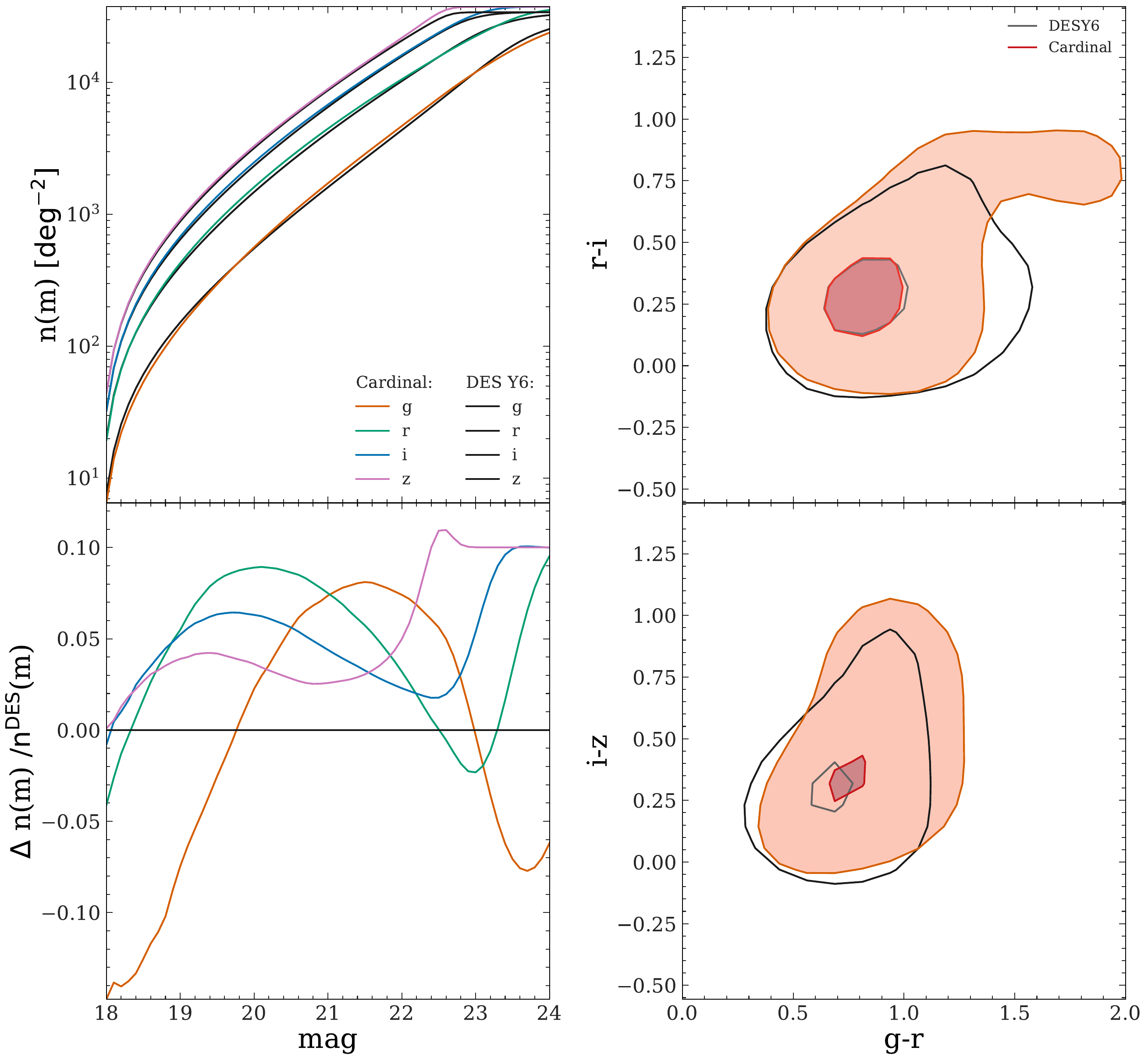}
    \caption{Comparison of apparent magnitudes in Cardinal and DES-Y6 \citep{Y6Gold}. Top left: over distribution. Bottom left: fraction differences. Top right: $g-r$ vs $r-i$ distributions. Bottom right: $i-z$ and $g-r$ distributions.}
    \label{app:magcomp}
\end{figure*}

We compare the magnitude distribution in Cardinal and DES-Y6 gold catalog in figure \ref{app:magcomp}. 
\section{Non-linear bias of halos}
\begin{figure}
    \centering
    \includegraphics[width=0.9\textwidth]{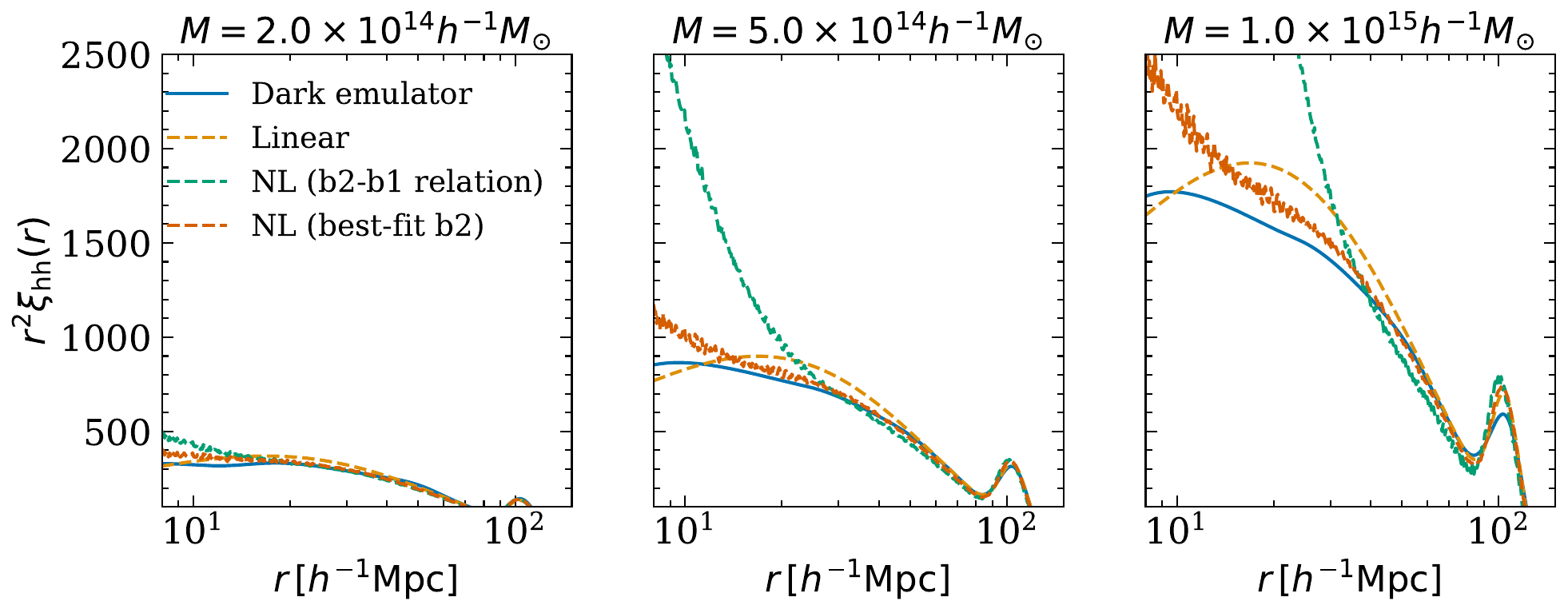}
    \caption{Comparisons of halo auto-correlation for halos with mass $2\times10^{14}\ h^{-1}M_{\odot}$ (left), $5\times10^{14}\ h^{-1}M_{\odot}$ (middle), $1\times10^{15}\ h^{-1}M_{\odot}$(right) at $z=0.5$. Blue lines show the prediction of the dark emulator, and orange dashed lines show the prediction assuming that halo clustering is linearly related to matter clustering. Green dashed lines and red dashed lines show prediction assuming a nonlinear biases model. Green lines assume $b_2$ values following the $b_2$--$b_1$ relation, while red lines use the best-fit value of $b_2$ obtained from comparing predictions of the theory and of the dark emulator.}
    \label{app:Darkemulator}
\end{figure}

In figure \ref{app:Darkemulator}, we compare our linear bias and non-linear bias model of halo clustering to predictions of Dark Emulator \citep{2019ApJ...884...29N}. The non-linear bias model is calculated using $b_2$--$b_1$ relation \citep{2016JCAP...02..018L}. As expected, the non-linear bias model improves the agreement at scales of $20-50\ h^{-1}{\rm Mpc}$ for halos with mass $=2\times10^{14}\ h^{-1}M_\odot$. However, at scales below $20\ h^{-1}{\rm Mpc}$, the non-linear bias model becomes worse than the linear bias model on agreements with the Dark Emulator. This is problematic because our scale cut for cluster clustering and cluster--galaxy cross-correlation are  $16\ h^{-1}{\rm Mpc}$ and $8\ h^{-1}{\rm Mpc}$. Adopting the non-linear bias model with $b_2$--$b_1$ relation will lead to an overly conservative conclusion. To solve this problem, we explore a model where the $b_2$ is fitted to the simulation to understand whether the model has the flexibility to fit the halo clustering. As shown in figure \ref{app:Darkemulator}, this model improves the agreement to the Dark Emulator to a much smaller scale compared to the $b_2$--$b_1$ relation model.

\begin{figure*}
    \centering
    \includegraphics[width=0.9\textwidth]{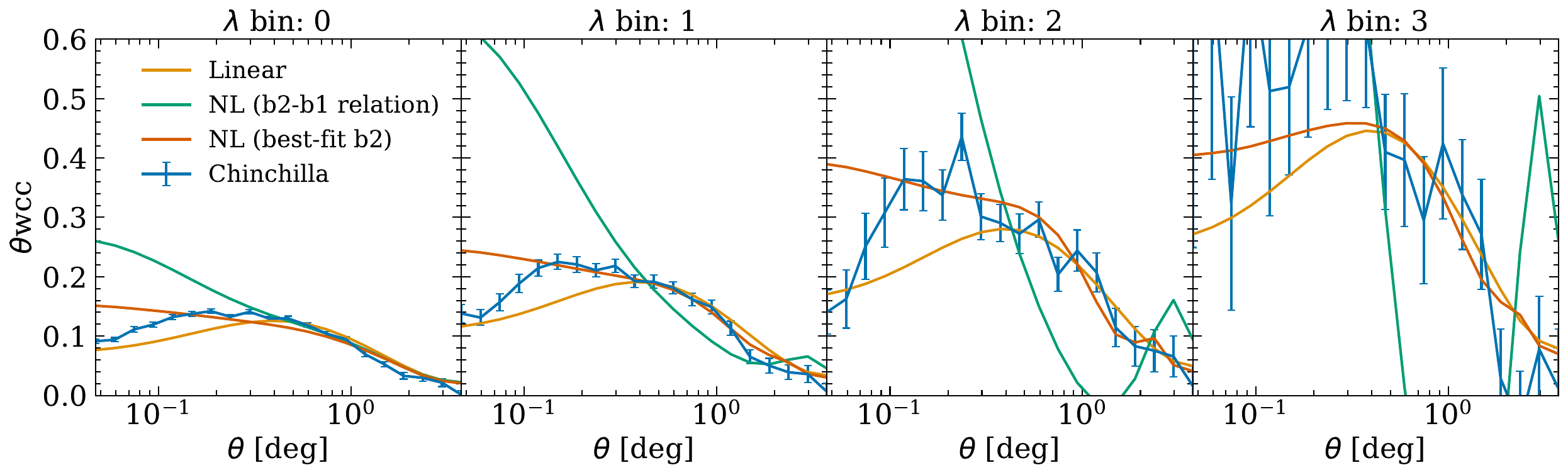}
    \caption{Comparisons of cluster clustering between measurements in Chinchilla N-body simulations and theory predictions for clusters in the highest redshift bins of the analysis. Four panels correspond to the four richness bins adopted in the analyses. Blue lines show the prediction of the Chinchilla simulations with error bars estimated from 100 Jackknife subsampling. Orange, green, and red lines show theory predictions. The orange line corresponds to the linear bias model. Green and red lines show predictions corresponding to the non-linear bias model. The $b_2$ value in the calculation of green lines comes from a $b_2$--$b_1$ relation, while the $b_2$ value for red lines comes from the best-fit value of the theory calculation when comparing to Chinchilla halos.}
    \label{app:NLChinchilla}
\end{figure*}

To further investigate this effect on our data vector, we employ halos in six $10k$ $\rm{deg}^2$ of Chinchilla N-body lightcone simulation \citep{JoeBuzzard} to compute the projected halo clustering. We adopt the same redshift cut and mass distribution as the fiducial model in the simulated likelihood analyses. In figure \ref{app:NLChinchilla}, we compare the linear and non-linear bias models in the highest redshift bin. Consistently to what we found in the $3D$ halo clustering, the projected halo clustering shows a similar conclusion: the non-linear model with $b_2$--$b_1$ relation disagrees with the N-body predictions more than the linear model. In comparison, the non-linear model with free $b_2$ has the flexibility to fit the N-body simulation and extend the model's fidelity to a smaller scale than the linear model. We then adopt the $b_2$ value fitted to the Chinchilla simulation to perform the simulated likelihood analysis. 
\label{app:NL}

\section{Selection bias model}
\label{app:HOD-cylindermock}
\begin{figure*}
    \centering
    \includegraphics[width=0.9\textwidth]{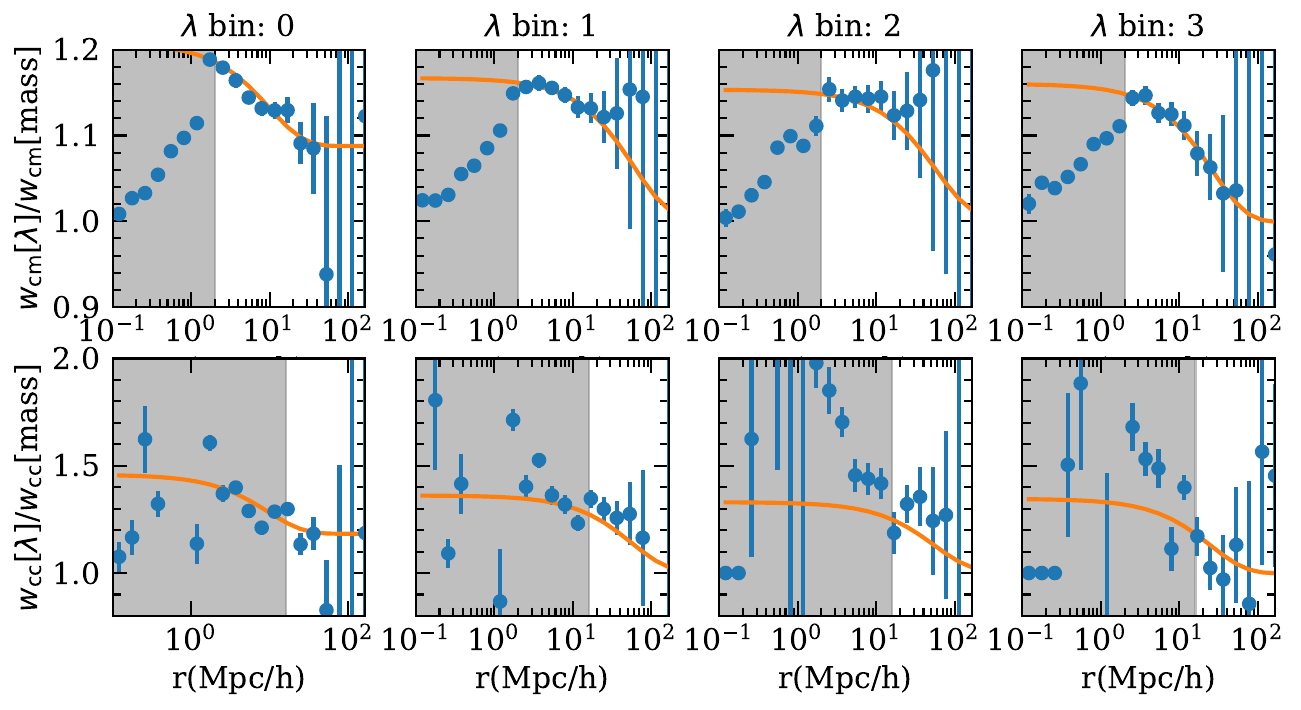}
    \caption{Ratios of cluster--matter (top) and cluster--cluster (bottom) correlations of richness-selected halos and randomly-selected halos that have the same mass distribution. Different panels correspond to different richness bins defined in section \ref{sec:mockuniverse}.  Dots show the measurement based on the cylinder mocks constructed from the Chinchilla N-body simulations. Error bars show $1\sigma$ uncertainties estimated from $100$ Jackknife resampling. Lines correspond to the best-fit selection bias model adopted in this analysis.}
    \label{app:selection}
\end{figure*}

\begin{figure*}
    \centering
    \includegraphics[width=0.9\textwidth]{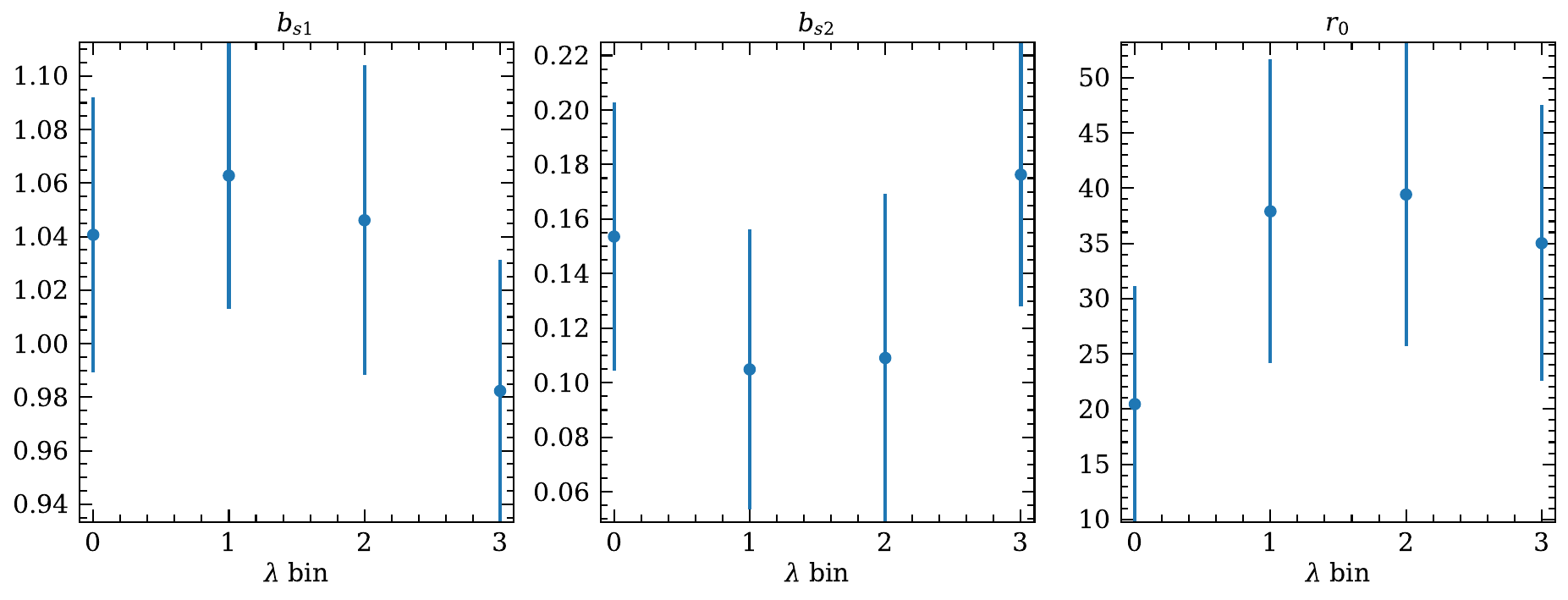}
    \caption{Best-fit values and their $1\sigma$ uncertainties of three parameters in the selection bias model. The values obtained from fitting to the cluster--matter correlations measured in the cylinder mocks constructed from the Chinchilla N-body simulations. Here we fit different richness bins individually and compare their values in each panel. We find no evident richness dependence of the parameter; thus, we adopt a richness-independent model in the analysis. Here, we show results in the highest redshift bin for clarity, but note that all redshift bin shows similar results.}
    \label{app:selection_2}
\end{figure*}

We generate HOD cylinder mocks following methods described in \citep{Heidiselection}. We check the length of the cylinder for 90, 120, and 150 $h^{-1}\rm{Mpc}$ and find the result is insensitive to those choices. We measure cluster--matter cross-correlation functions using the cylinder-selected clusters and compare that with the randomly selected halos that have the same halo mass distribution.  The ratio of the two gives us an empirical estimation of the selection bias effect on cluster density fields. We fit our model described in section \ref{sec:selection_model} to these measurements and find good agreement (figure \ref{app:selection}). As an additional check, we also measure the clustering of clusters in these simulations and compare that with our model predictions that fit the cluster--matter cross-correlations (see lower panels). We also find good agreement. Finally, in figure \ref{app:selection_2}, we check the richness dependence of the parameters in this mock and find no detection. We, therefore, adopt a richness-independent model in our fiducial analyses.  

\section{Hmcode2020 implementation in CLASS }
\begin{figure*}
    \centering
    \includegraphics[width=0.9\textwidth]{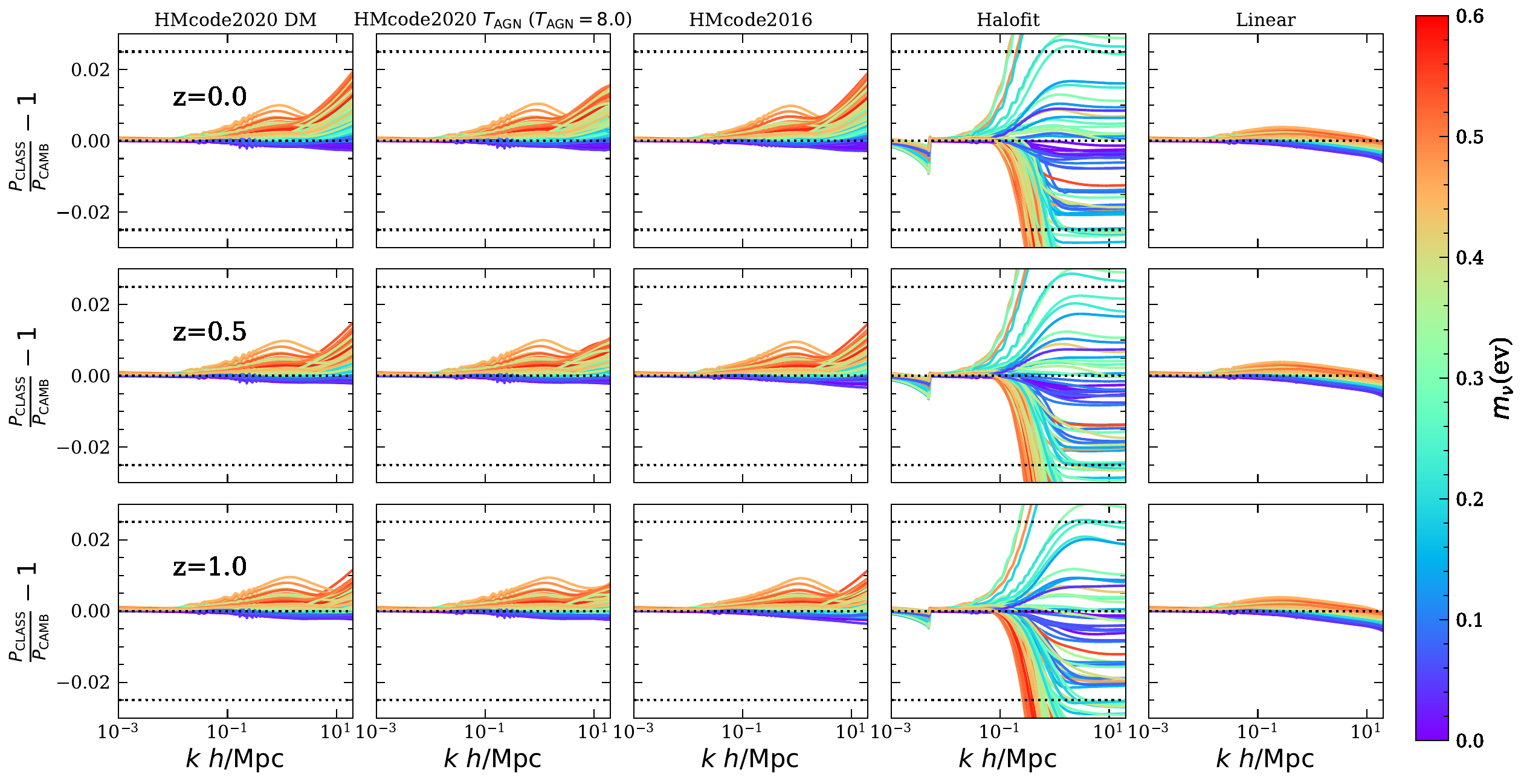}
    \caption{Comparison of theory calculations in CAMB and CLASS. Each line corresponds to a random cosmology drawn from the prior in Table \ref{tab:params}, color-coded by the neutrino mass. Different rows correspond to a different redshift, while different columns correspond to different models of power spectra. The first column shows the comparison of HMcode2020. The second column shows HMcode2020 with baryons, where we fix $T_{\rm AGN}$ to $8.0$. We verify that the plot does not change with a different $T_{\rm AGN}$ value. The third to fifth columns show comparisons of the existing implementation in CAMB and CLASS: Hmcode-2016 \citep{hmcode2016}, Halofit \citep{Takahashi2012}, and the linear calculation.}
    \label{app:camvsclass}
\end{figure*}

We implement HMcode2020 \citep{HMcode2020} model in CLASS v2.9.3 \citep{CLASS}. We follow closely with the HMcode2020 implementation in CAMB but with one important difference in the treatment of massive neutrinos. In the calculation of the variance in the linear theory of cold matter and baryon fluid ($\sigma(R)$), filtered on a Lagrangian scale $R$, one needs to evaluate the linear power spectra of cold dark matter and baryons. In the CAMB implementation, this is calculated with Eisenstein--Hu fitting function \citep{1998ApJ...496..605E}. In our implementation, we follow the implementation of HMcode2016 in CLASS, where $\sigma(R)$ is calculated from the Boltzmann solver. In figure \ref{app:camvsclass}, we compare different non-linear power spectra models in CAMB and CLASS. To do this, we generate 200 random samples from DES cosmology priors (table \ref{tab:params}). We color-coded the differences based on the sum of neutrino masses. The deviation between our implementation of HMcode2020 and the CAMB version is much larger than the differences in the linear theory calculation and is correlated with neutrino mass. However, this difference is smaller than the expected $2.5\%$ uncertainties of the HMcode2020 model. In the case of no massive neutrinos, we find that the fractional difference between our implemented HMcode2020 and HMcode2020 in CAMB is $\simeq 0.3\%$, much better than the case of massive neutrinos. Further, we note that the level of deviation between CAMB and CLASS in our implemented HMcode2020 is similar to that of the public HMcode2016 \citep{hmcode2016}. We suspect that this is due to the same reason of $\sigma(R)$ calculations, but investigating the source of differences in HMcode2016 is beyond the scope of this paper.

\section{Systematic bias in multiprobe analyses}
\label{app:biasmultiprobe}
In this section, we explore the impact of combining different probes on the ratio of systematic to statistical errors. Typically, analyses ensure that systematics contribute only a small fraction of the statistical uncertainties for individual probes (e.g., \ttt{}). However, it is less common to verify whether systematics remain subdominant when multiple uncorrelated probes are combined (e.g., \ttt{}, BAO, and SN). A common assumption is that systematics from different probes are uncorrelated, and thus combining them should pull constraints closer to the true values. Here, we demonstrate that this assumption does not always hold; the ratio of systematics to statistical errors can actually increase due to degeneracy breaking. This shows the importance of testing the impact of systematics on the final combined data vector.

\begin{figure*}
    \centering
    \includegraphics[width=0.9\textwidth]{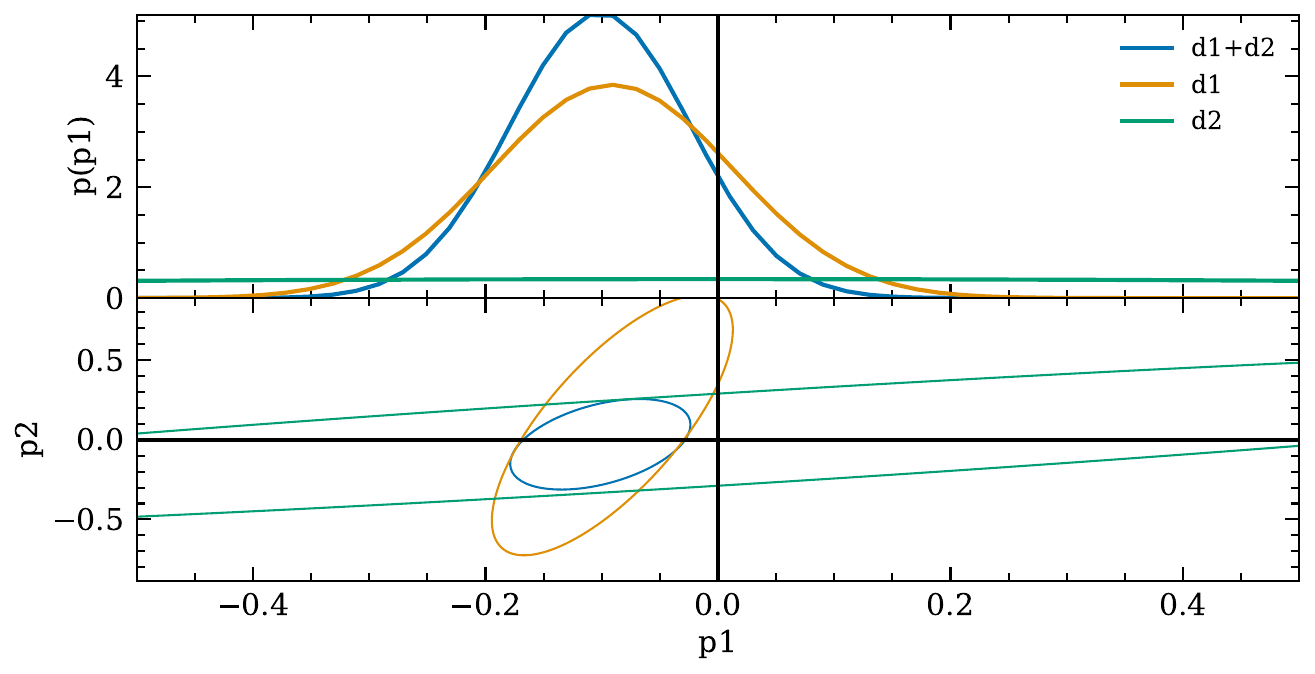}
    \caption{Demonstration that the ratio of systematic bias to statistical uncertainties could be enlarged when combined with independent and noise-free data vector due to degeneracy breaking.}
    \label{app:bias_demonstration}
\end{figure*}

To illustrate this, we consider two independent data vectors ($d_1$ and $d_2$), each containing two entries that depend on two parameters ($p_1$ and $p_2$). We assume no correlations between the two data vectors and introduce a systematic that biases the inference of $p_1$ in the first data vector but has no effect on the second. Using the Fisher formalism, we calculate the one-sigma parameter constraints and the systematic-induced bias in the parameter values. As shown in Fig.~\ref{app:bias_demonstration}, we find that while the systematics bias the $d_1$ analysis by $0.88\sigma$, they bias the combined $d_1$ and $d_2$ analysis by $1.3\sigma$.

\section{List of improvements from \citep{4x2pt1, 4x2pt2}}
\label{app:listofmodel}
We first list all model changes compared to the To$\&$Krause2021 model with brief justifications. 
\begin{enumerate}
    \item Non-local cluster lensing estimator: It has been shown that cluster lensing estimator $\gamma_t$ contains one-halo contribution at large scales due to the non-local nature of the estimator \citep{Ytransform}. 
    \item Intrinsic alignment: To be consistent with \ttt{} analysis choices. 
    \item Magnification due to line-of-sight structure: To be consistent with \ttt{} analysis choices. 
    \item Exact Legendre polynomial summation and proper bin average for covariance matrix calculation: Improve numerical stability of covariance matrix calculation. 
    \item Scale-dependent scale bias model: We find it necessary to adopt a scale-dependent selection bias model to meet the precision of DES-Y6. 
    \item We adopt HMCode2020 to model the non-linear matter power spectrum, which we implement in CLASS v2.9: To be consistent with \ttt{} analysis choice. 
    \item We adopt a new machine learning-based sampler\citep{linna}: To conduct the analysis in a reasonable amount of time.
    \item We adopt non-limber calculations when generating a covariance matrix: To explicitly calculate the covariance between samples in different tomographic bins.  
    \item We modify the calculation of non-linear halo bias parameters: We find a significant bias when comparing the calculation in \citep{4x2pt1} to simulations. 
\end{enumerate}

We then list improvements in the validation methods:
\begin{enumerate}
    \item We construct new simulations to test the model: see \citep{cardinal} for a comprehensive list of improvements.
    \item We check contaminations for \clustercomb{},  \allcomb{}, and \allcomb{}+BAO+SN.
    \item We consider a more comprehensive list of contaminations, including baryonic impacts on matter power spectra, the accuracy of the matter power spectrum model, and non-linear cluster lensing.
\end{enumerate}
\input{affiliations_super}

\end{document}

%% file: authorsuper.tex
\author{Chun-Hao To\textsuperscript{1}}
\author{Elisabeth Krause\textsuperscript{2}}
\author{Chihway Chang\textsuperscript{1,3}}
\author{Hao-Yi Wu\textsuperscript{4}}
\author{Risa H. Wechsler\textsuperscript{5,6,7}}
\author{Eduardo Rozo\textsuperscript{8}}
\author{David H. Weinberg\textsuperscript{9,10}}
\author{D.~Anbajagane\textsuperscript{3}}
\author{S.~Avila\textsuperscript{11}}
\author{J.~Blazek\textsuperscript{12}}
\author{S.~Bocquet\textsuperscript{13}}
\author{M.~Costanzi\textsuperscript{14,15,16}}
\author{J.~De~Vicente\textsuperscript{11}}
\author{J.~Elvin-Poole\textsuperscript{17}}
\author{A.~Fert\'e\textsuperscript{7}}
\author{S.~Grandis\textsuperscript{13}}
\author{J.~Muir\textsuperscript{18,19}}
\author{A.~Porredon\textsuperscript{11,20}}
\author{S.~Samuroff\textsuperscript{12,21}}
\author{E.~Sanchez\textsuperscript{11}}
\author{D.~Sanchez Cid\textsuperscript{11,22}}
\author{I.~Sevilla-Noarbe\textsuperscript{11}}
\author{N.~Weaverdyck\textsuperscript{23,24}}
\author{T.~M.~C.~Abbott\textsuperscript{25}}
\author{M.~Aguena\textsuperscript{26}}
\author{F.~Andrade-Oliveira\textsuperscript{22}}
\author{D.~Bacon\textsuperscript{27}}
\author{M.~R.~Becker\textsuperscript{28}}
\author{D.~Brooks\textsuperscript{29}}
\author{A.~Carnero~Rosell\textsuperscript{30,26,31}}
\author{J.~Carretero\textsuperscript{21}}
\author{A.~Choi\textsuperscript{32}}
\author{L.~N.~da Costa\textsuperscript{26}}
\author{M.~E.~S.~Pereira\textsuperscript{33}}
\author{T.~M.~Davis\textsuperscript{34}}
\author{S.~Desai\textsuperscript{35}}
\author{P.~Doel\textsuperscript{29}}
\author{S.~Everett\textsuperscript{36}}
\author{J.~Frieman\textsuperscript{1,37,3}}
\author{J.~Garc\'ia-Bellido\textsuperscript{38}}
\author{M.~Gatti\textsuperscript{3}}
\author{E.~Gaztanaga\textsuperscript{39,27,40}}
\author{G.~Giannini\textsuperscript{21,3}}
\author{D.~Gruen\textsuperscript{13}}
\author{G.~Gutierrez\textsuperscript{37}}
\author{S.~R.~Hinton\textsuperscript{34}}
\author{D.~L.~Hollowood\textsuperscript{41}}
\author{K.~Honscheid\textsuperscript{9,42}}
\author{T.~Jeltema\textsuperscript{41}}
\author{K.~Kuehn\textsuperscript{43,44}}
\author{S.~Lee\textsuperscript{45}}
\author{J.~L.~Marshall\textsuperscript{46}}
\author{J. Mena-Fern\textsuperscript{47}}
\author{R.~Miquel\textsuperscript{48,21}}
\author{J.~J.~Mohr\textsuperscript{49,50}}
\author{J.~Myles\textsuperscript{51}}
\author{A.~Palmese\textsuperscript{52}}
\author{A.~A.~Plazas~Malag\'on\textsuperscript{6,7}}
\author{A.~K.~Romer\textsuperscript{53}}
\author{T.~Shin\textsuperscript{54}}
\author{M.~Smith\textsuperscript{55}}
\author{E.~Suchyta\textsuperscript{56}}
\author{G.~Tarle\textsuperscript{57}}
\author{V.~Vikram\textsuperscript{}}
\author{A.~R.~Walker\textsuperscript{25}}
\author{J.~Weller\textsuperscript{50,58}}

%% file: affiliations_super.tex
\section*{Affiliations}
\noindent
\textsuperscript{1} Department of Astronomy and Astrophysics, University of Chicago, Chicago, IL 60637, USA\\
\textsuperscript{2} Department of Astronomy/Steward Observatory, University of Arizona, 933 North Cherry Avenue, Tucson, AZ 85721-0065, USA\\
\textsuperscript{3} Kavli Institute for Cosmological Physics, University of Chicago, Chicago, IL 60637, USA\\
\textsuperscript{4} Department of Physics, Southern Methodist University, Dallas, TX 75205, USA\\
\textsuperscript{5} Department of Physics, Stanford University, 382 Via Pueblo Mall, Stanford, CA 94305, USA\\
\textsuperscript{6} Kavli Institute for Particle Astrophysics \& Cosmology, P. O. Box 2450, Stanford University, Stanford, CA 94305, USA\\
\textsuperscript{7} SLAC National Accelerator Laboratory, Menlo Park, CA 94025, USA\\
\textsuperscript{8} Department of Physics, University of Arizona, Tucson, AZ 85721, USA\\
\textsuperscript{9} Center for Cosmology and Astro-Particle Physics, The Ohio State University, Columbus, OH 43210, USA\\
\textsuperscript{10} Department of Astronomy, Ohio State University, Columbus, OH 43210, USA\\
\textsuperscript{11} Centro de Investigaciones Energ\'eticas, Medioambientales y Tecnol\'ogicas (CIEMAT), Madrid, Spain\\
\textsuperscript{12} Department of Physics, Northeastern University, Boston, MA 02115, USA\\
\textsuperscript{13} University Observatory, Faculty of Physics, Ludwig-Maximilians-Universit\"at, Scheinerstr. 1, 81679 Munich, Germany\\
\textsuperscript{14} Astronomy Unit, Department of Physics, University of Trieste, via Tiepolo 11, I-34131 Trieste, Italy\\
\textsuperscript{15} INAF-Osservatorio Astronomico di Trieste, via G. B. Tiepolo 11, I-34143 Trieste, Italy\\
\textsuperscript{16} Institute for Fundamental Physics of the Universe, Via Beirut 2, 34014 Trieste, Italy\\
\textsuperscript{17} Department of Physics and Astronomy, University of Waterloo, 200 University Ave W, Waterloo, ON N2L 3G1, Canada\\
\textsuperscript{18} Department of Physics, University of Cincinnati, Cincinnati, Ohio 45221, USA\\
\textsuperscript{19} Perimeter Institute for Theoretical Physics, 31 Caroline St. North, Waterloo, ON N2L 2Y5, Canada\\
\textsuperscript{20} Ruhr University Bochum, Faculty of Physics and Astronomy, Astronomical Institute, German Centre for Cosmological Lensing, 44780 Bochum, Germany\\
\textsuperscript{21} Institut de F\'\\
\textsuperscript{22} Physik-Institut, University of Zürich, Winterthurerstrasse 190, CH-8057 Zürich, Switzerland\\
\textsuperscript{23} Department of Astronomy, University of California, Berkeley,  501 Campbell Hall, Berkeley, CA 94720, USA\\
\textsuperscript{24} Lawrence Berkeley National Laboratory, 1 Cyclotron Road, Berkeley, CA 94720, USA\\
\textsuperscript{25} Cerro Tololo Inter-American Observatory, NSF's National Optical-Infrared Astronomy Research Laboratory, Casilla 603, La Serena, Chile\\
\textsuperscript{26} Laborat\'orio Interinstitucional de e-Astronomia - LIneA, Av. Pastor Martin Luther King Jr, 126 Del Castilho, Nova Am\'erica Offices, Torre 3000/sala 817 CEP: 20765-000, Brazil\\
\textsuperscript{27} Institute of Cosmology and Gravitation, University of Portsmouth, Portsmouth, PO1 3FX, UK\\
\textsuperscript{28} Argonne National Laboratory, 9700 South Cass Avenue, Lemont, IL 60439, USA\\
\textsuperscript{29} Department of Physics \& Astronomy, University College London, Gower Street, London, WC1E 6BT, UK\\
\textsuperscript{30} Instituto de Astrofisica de Canarias, E-38205 La Laguna, Tenerife, Spain\\
\textsuperscript{31} Universidad de La Laguna, Dpto. Astrofísica, E-38206 La Laguna, Tenerife, Spain\\
\textsuperscript{32} NASA Goddard Space Flight Center, 8800 Greenbelt Rd, Greenbelt, MD 20771, USA\\
\textsuperscript{33} Hamburger Sternwarte, Universit\"\\
\textsuperscript{34} School of Mathematics and Physics, University of Queensland,  Brisbane, QLD 4072, Australia\\
\textsuperscript{35} Department of Physics, IIT Hyderabad, Kandi, Telangana 502285, India\\
\textsuperscript{36} California Institute of Technology, 1200 East California Blvd, MC 249-17, Pasadena, CA 91125, USA\\
\textsuperscript{37} Fermi National Accelerator Laboratory, P. O. Box 500, Batavia, IL 60510, USA\\
\textsuperscript{38} Instituto de Fisica Teorica UAM/CSIC, Universidad Autonoma de Madrid, 28049 Madrid, Spain\\
\textsuperscript{39} Institut d'Estudis Espacials de Catalunya (IEEC), 08034 Barcelona, Spain\\
\textsuperscript{40} Institute of Space Sciences (ICE, CSIC),  Campus UAB, Carrer de Can Magrans, s/n,  08193 Barcelona, Spain\\
\textsuperscript{41} Santa Cruz Institute for Particle Physics, Santa Cruz, CA 95064, USA\\
\textsuperscript{42} Department of Physics, The Ohio State University, Columbus, OH 43210, USA\\
\textsuperscript{43} Australian Astronomical Optics, Macquarie University, North Ryde, NSW 2113, Australia\\
\textsuperscript{44} Lowell Observatory, 1400 Mars Hill Rd, Flagstaff, AZ 86001, USA\\
\textsuperscript{45} Jet Propulsion Laboratory, California Institute of Technology, 4800 Oak Grove Dr., Pasadena, CA 91109, USA\\
\textsuperscript{46} George P. and Cynthia Woods Mitchell Institute for Fundamental Physics and Astronomy, and Department of Physics and Astronomy, Texas A\&M University, College Station, TX 77843,  USA\\
\textsuperscript{47} LPSC Grenoble - 53, Avenue des Martyrs 38026 Grenoble, France\\
\textsuperscript{48} Instituci\'o Catalana de Recerca i Estudis Avan\c\\
\textsuperscript{49} Faculty of Physics, Ludwig-Maximilians-Universit\"at, Scheinerstr. 1, 81679 Munich, Germany\\
\textsuperscript{50} Max Planck Institute for Extraterrestrial Physics, Giessenbachstrasse, 85748 Garching, Germany\\
\textsuperscript{51} Department of Astrophysical Sciences, Princeton University, Peyton Hall, Princeton, NJ 08544, USA\\
\textsuperscript{52} Department of Physics, Carnegie Mellon University, Pittsburgh, Pennsylvania 15312, USA\\
\textsuperscript{53} Department of Physics and Astronomy, Pevensey Building, University of Sussex, Brighton, BN1 9QH, UK\\
\textsuperscript{54} Department of Physics and Astronomy, Stony Brook University, Stony Brook, NY 11794, USA\\
\textsuperscript{55} Physics Department, Lancaster University, Lancaster, LA1 4YB, UK\\
\textsuperscript{56} Computer Science and Mathematics Division, Oak Ridge National Laboratory, Oak Ridge, TN 37831\\
\textsuperscript{57} Department of Physics, University of Michigan, Ann Arbor, MI 48109, USA\\
\textsuperscript{58} Universit\"ats-Sternwarte, Fakult\"at f\"ur Physik, Ludwig-Maximilians Universit\"at M\"unchen, Scheinerstr. 1, 81679 M\"unchen, Germany